\documentclass[12pt,aps,nofootinbib,pra]{revtex4-1}
\usepackage{amsfonts,amsmath,amssymb,amsthm}
\usepackage{acronym,array,bm,ccaption,color,dcolumn,epsfig,graphicx,nomencl,revsymb4-1,upgreek,url}
\usepackage[pdftex]{hyperref} 
\hypersetup{colorlinks=true, pdfauthor={Matthew D. Grace, et al.},
pdftitle={Optimized pulses for the control of uncertain qubits}}

\newcommand{\Cbb}{\mathbb{C}}

\newcommand{\Cfrak}{\mathfrak{C}}

\newcommand{\Fcal}{\mathcal{F}}

\newcommand{\Hcal}{\mathcal{H}}

\newcommand{\Hfrak}{\mathfrak{H}}

\newcommand{\Imag}{\mathrm{Im}}

\newcommand{\Jcal}{\mathcal{J}}

\newcommand{\Kcal}{\mathcal{K}}

\newcommand{\Lcal}{\mathcal{L}}

\newcommand{\Mcal}{\mathcal{M}}

\newcommand{\Qbb}{\mathbb{Q}}

\newcommand{\Rbb}{\mathbb{R}}
\newcommand{\Rcal}{\mathcal{R}}

\newcommand{\rHS}{\mathrm{HS}}
\newcommand{\rmB}{\mathrm{B}}

\newcommand{\rmd}{\mathrm{d}}
\newcommand{\rme}{\mathrm{e}}

\newcommand{\rmh}{\mathrm{h}}

\newcommand{\rmi}{\mathrm{i}}

\newcommand{\rmo}{\mathrm{o}}

\newcommand{\rmr}{\mathrm{r}}

\newcommand{\rmT}{\mathrm{T}}

\newcommand{\rmu}{\mathrm{u}}

\newcommand{\Scal}{\mathcal{S}}
\newcommand{\sign}{\mathrm{sgn}}

\newcommand{\SUrm}{\mathrm{SU}}
\newcommand{\surm}{\mathrm{su}}

\newcommand{\tf}{t_{\mathrm{f}}}

\newcommand{\Tr}{\mathrm{Tr}}

\newcommand{\Urm}{\mathrm{U}}

\newcommand{\Utf}{U_{\tf}}

\newcommand{\Xbb}{\mathbb{X}}

\begin{document}

\title{Optimized pulses for the control of uncertain qubits}

\author{Matthew D.~Grace}
\email[Electronic address: ]{mgrace@sandia.gov}
\affiliation{Department of Scalable \& Secure Systems Research,
 Sandia National Laboratories, Livermore, CA 94550}
\author{Jason M.~Dominy\footnote{Current address: Center for Quantum
    Information Science \& Technology, University of Southern
    California, Los Angeles, CA 90089}}
\email[Electronic address: ]{jdominy@usc.edu}
\affiliation{Program in Applied \& Computational Mathematics, Princeton
 University, Princeton, New Jersey 08544}
\author{Wayne M.~Witzel}
\email[Electronic address: ]{wwitzel@sandia.gov}
\affiliation{Department of Advanced Device Technologies,
 Sandia National Laboratories, Albuquerque, NM 87185}
\author{Malcolm S.~Carroll}
\email[Electronic address: ]{mscarro@sandia.gov}
\affiliation{Department of Photonic Microsystem Technologies,
 Sandia National Laboratories, Albuquerque, NM 87185}

\date{\today}

\begin{abstract}
Constructing high-fidelity control fields that are robust to control,
system, and/or surrounding environment uncertainties is a crucial
objective for quantum information processing. Using the two-state
Landau-Zener model for illustrative simulations of a controlled qubit,
we generate optimal controls for $\pi/2$- and $\pi$-pulses, and
investigate their inherent robustness to uncertainty in the magnitude of
the drift Hamiltonian. Next, we construct a quantum-control protocol to
improve system-drift robustness by combining environment-decoupling
pulse criteria and optimal control theory for unitary operations. By
perturbatively expanding the unitary time-evolution operator for an open
quantum system, previous analysis of environment-decoupling control
pulses has calculated explicit control-field criteria to suppress 
environment-induced errors up to (but not including) third order from
$\pi/2$- and $\pi$-pulses. We systematically integrate this criteria
with optimal control theory, incorporating an estimate of the uncertain
parameter, to produce improvements in gate fidelity \emph{and}
robustness, demonstrated via a numerical example based on double quantum
dot qubits. For the qubit model used in this work, \emph{post facto}
analysis of the resulting controls suggests that realistic control-field
fluctuations and noise may contribute just as significantly to gate
errors as system and environment fluctuations.
\end{abstract}


\maketitle

\section{Introduction}
\label{sec:introduction}

Demanding requirements for gate fidelities to achieve
fault-tolerant quantum computation (QC) \cite{Gaitan08a} have motivated
the need for improved quantum control protocols (QCPs). In quantum
information science \cite{Ladd10a}, there are (at least) three distinct
dynamical approaches to improving the fidelity of qubit operations in
the presence of environmental interactions: dynamical-decoupling (DD)
pulse sequences \cite{Liu07a, Uhrig08a}, optimal control theory (OCT)
\cite{Roloff09a, Brif10a}, and quantum error-correcting codes (QECCs)
\cite{Gaitan08a}. Although interesting non-dynamical methods exist for
noise suppression, e.g., decoherence-free subspaces (DFSs)
\cite{Lidar03a}, noiseless subsystems \cite{Zanardi97a}, and
anyonic/topological systems \cite{Nayak08a}, the work reported in this
article focuses exclusively on dynamical approaches for controlling
quantum systems \cite{Werschnik07a}. Specifically, the objective of this
work is to construct a \emph{hybrid} QCP, combining methods and results
from DD and OCT, to locate control fields that (a) produce high-fidelity
rotations about a particular axis \emph{and} (b) are robust with respect
to an uncertain frequency of rotation about an orthogonal axis. By
combining complementary features of these analytically- and
numerically-based QCPs, we have developed a hybrid QCP where estimates
of system parameters can be directly incorporated into numerical
simulations to generate improved quantum operations.

During the past few years, P.~Karbach, S.~Pasini, G.~Uhrig, and
colleagues have made significant contributions toward the mathematical
analysis and design of DD pulses and sequences for controlling qubit
systems and decoupling them from their surrounding environment, e.g.,
refs.~\cite{Uhrig07a, Uhrig08a, Karbach08a, Pasini08a, Pasini08b,
Pasini09a, Pasini10a, Pasini11a}. In a recent article \cite{Pasini09a},
using a rather general open-system, time-dependent Hamiltonian for one
qubit, they derive analytical control-field criteria for $\pi/2$- and
$\pi$-pulses, which, when satisfied, eliminate the first- and
second-order errors in the unitary time-evolution operator resulting
from qubit-environment interactions. In this work, we refer to these
criteria as ``decoupling-pulse criteria'' (DPC) and to the control
fields that satisfy this criteria as ``decoupling pulses'' (DPs). We
adapt the DPC for the case of closed-system unitary control, where the
dynamics are influenced by an uncertain drift (i.e., time-independent)
term in the qubit Hamiltonian. For control fields that satisfy the DPC,
this adaptation eliminates the first- and second-order effects resulting
from the drift term. Using a novel method for multi-objective control,
we combine the mathematical DPC with a numerical procedure based on OCT
for unitary control \cite{Palao03a} that incorporates an estimate of the
drift-term magnitude (i.e., system information) to construct control
fields with increased fidelity and robustness to uncertainty in the
drift term. For brevity, we refer to this combination of the DPC and OCT
as ``DPC+OCT''. To demonstrate the utility of our approach, we optimize
and evaluate these control fields using a qubit model based on the
two-level Landau-Zener Hamiltonian \cite{Messiah99a} that has an
uncertain drift term and is driven by a deterministic control field.
Even though the qubit model is quite general, i.e., the Hamiltonian
employed represents a one-qubit system with a linear drift term, driven
by a scalar control field, and thus describes a variety of qubits (e.g.,
atomic, spin, superconducting, etc.~\cite{Ladd10a}), we select physical
units for the model that are relevant to double quantum dot (DQD) qubits
to investigate the practical features of our results \cite{Levy11a}.
With this model system, we demonstrate that the DPC+OCT combination can
be used to produce (a) improved fidelity compared to DPs alone and (b)
improved robustness to uncertainty in the drift magnitude compared to
results from DPs and OCT alone. Although research examining the effects
of classical control noise is \emph{extremely} important for practical
QC \cite{vanEnk02a, Sola04a}, all control fields in this work are
assumed to be deterministic, with control amplitudes that are exact to
numerical precision. Thus, uncertainty is assumed to be present only in
the system, and, unless otherwise specified, control robustness refers
to robustness to drift uncertainty.

Related research on hybrid QCPs includes that performed by Lidar
\emph{et al.}, who proposed the application of DD pulse sequences on
\emph{logical} qubits encoded in DFSs or QECCs to eliminate decoherence
in solid-state and trapped-ion qubits \cite{Byrd02b, Byrd03a,
Khodjasteh03a, Ng11a}. Borneman \emph{et al.} used OCT to design
control fields that are robust to systematic amplitude and resonance
inhomogeneities, thereby improving the performance of the so-called 
Carr-Purcell-Meiboom-Gill pulse sequence \cite{Borneman10a}. In
addition, there have been many studies on the control and
controllability of (inhomogeneous) quantum-mechanical ensembles, such as
a collection of coupled or uncoupled spin systems, primarily for
state-based objectives, e.g., refs.~\cite{Khaneja05a, Li06b, Dong09a,
Ruths10a, Ruths11a}, and sequences of unitary time-evolution operators
that compensate for systematic off-resonant effects, e.g.,
\cite{Cummins00a, Cummins03a, Brown04a, Alway07a, Jones11a}.

This article is organized as follows. Section \ref{sec:LZ-model}
introduces and develops the model qubit system based on the Landau-Zener
Hamiltonian \cite{Messiah99a} used in our optimizations and simulations.
For illustrative purposes, this model is compared to a logical DQD
semiconductor qubit, where uncertainty in the drift term of the system
Hamiltonian is due to the surrounding nuclear spin environment 
\cite{Witzel06a, Witzel07a}. A system of scaled units is defined that
allows for the comparison of our model and control fields to relevant
experimental parameters. In section \ref{sec:control-GrA}, we summarize
our gradient-based OCT routine for deterministic Hamiltonian systems,
describing our objective functional for unitary control, relevant
control properties, and the numerical optimization procedure. Section
\ref{sec:results-OCT} presents results from unitary OCT for subsequent
comparison to those from the DPC and DPC+OCT QCPs. Inherent robustness
of these optimal controls (OCs) to variations in the magnitude of the
uncertain drift term is also analyzed, and a functional is proposed to
quantify this robustness. For the individual unitary targets considered,
despite the similar structures of the resulting OCs for different
drift-term magnitudes, their gate distances as a function of the drift
magnitude differ dramatically. Section \ref{sec:robust-DPC} summarizes
the nonlinear control-field criteria developed by Pasini \emph{et al.}
for designing control pulses that are robust to decoherence
\cite{Pasini09a}. Our adaptation of these criteria to closed-system
unitary control is explained and the hybrid DPC+OCT control problem is
posed. Control fields satisfying this criteria are applied to our model
qubit system and their robustness is analyzed. In section
\ref{sec:DPC+OCT}, we describe and mathematically formulate our
gradient-based method for solving the nonlinearly-constrained control 
problem. Section \ref{sec:results-DPC+OCT} presents results from our
DPC+OCT optimization algorithm. Gate distance and robustness of the
control fields are numerically analyzed and discussed. We also compare
OCT, DPC, and DPC+OCT results collectively to illustrate the benefit of
our hybrid approach. In addition to comparing the gate distances
directly, we apply these controls to an inhomogeneous ensemble of
systems to emphasize the improvement that may be obtained from this
hybrid QCP. Like the OCT results, significant gate-distance
sensitivities to relatively small control-field differences are observed
for the DPC+OCT controls, supporting further study and suppression of
the effects of undesired control-field fluctuations and noise on quantum
information processing. We conclude this article in section
\ref{sec:conclusion} with a summary of our results and identify several 
future directions of our research.

\section{Landau-Zener model system}
\label{sec:LZ-model}

\subsection{Model Hamiltonian}

We represent the dynamical model of a qubit with the following
Hamiltonian (where $\hbar = 1$; details regarding units appear in
section \ref{ssec:units}):
\begin{equation}
\label{eq:Hamiltonian}
H(t) := \varepsilon S_{x} + C(t) S_{z},
\end{equation}
where $S_{\lambda} := \sigma_{\lambda}/2$ is a spin operator for a
spin-$1/2$ particle, $\sigma_{\lambda}$ is a Pauli matrix $\left(
\lambda \in \{x, y, z\} \right)$, $\varepsilon S_{x}$ represents a
persistent rotation about the $x$-axis, and $C(t)$ represents the
time-dependent control field driving rotations about the $z$-axis. Note
that $H(t)$ corresponds to the two-state Landau-Zener model
\cite{Messiah99a}, and that both $C(t) / \hbar$ and $\varepsilon /
\hbar$ have units of angular frequency, e.g., radians per second in SI
units.

Let $\Hcal$ denote the Hilbert space of the system, where $n := \dim\{
\Hcal \}$ ($n = 2$ for one qubit), and $\left\{ |\Scal_{i}\rangle
\right\}$ denote the orthonormal basis of $S_{z}$ that spans $\Hcal$,
with corresponding eigenvalues $\pm 1/2$. The Lie group of all unitary
operators on $\Hcal$ is denoted by $\Urm(\Hcal)$. In general, the
unitary time-evolution operator $U(t) \in \Urm(\Hcal)$ for a closed
quantum system obeys the time-dependent Schr\"{o}dinger equation:
\begin{equation} 
\label{eq:Schrodinger}
\dot{U}(t) = -\rmi H(t)U(t),
\end{equation}
where $U(t = 0) = \openone_{n}$, the $n \times n$ identity matrix. From
a controllability perspective \cite{Huang83a, Rama95a}, the Hamiltonian
in eq.~\eqref{eq:Hamiltonian} generates the Lie algebra $\surm(2)$.
Thus, the system is completely dynamically controllable, i.e., any
element of the Lie group $\SUrm(2)$ can be generated via
eq.~\eqref{eq:Schrodinger} and an appropriately-shaped control
field. However, this analysis does not necessarily reveal anything about
the control-field structure required to realize an arbitrary $\SUrm(2)$
operation. As an illustrative example of our DPC+OCT QCP, we focus on
constructing unitary operations corresponding to $\pi/2$- and
$\pi$-rotations about the $z$-axis.

\subsection{Double quantum dot logical qubit}
\label{ssec:DQD-qubit}

Although our qubit model is quite general, for illustrative purposes we
refer to a particular application of a DQD solid-state qubit
\cite{Hanson07a, Taylor07a}, which has been studied in an array of
experiments, e.g., \cite{Johnson05a, Petta05a, Petta08a, Foletti09a,
Bluhm10a}. With one electron in each quantum dot, the DQD system spans
four spin-1/2 states. An applied magnetic field will break the
degeneracy of the states in which both electrons are either aligned
against or with the field. In this situation, it is possible (and often
advantageous) to work within the two-level subspace where the net spin
angular momentum is zero. By adjusting voltages in the electrostatically
defined quantum dots, the magnitude of the exchange interaction between
the electrons may be controlled. This interaction controls the splitting
between the singlet, $\lvert S \rangle := (\lvert \uparrow \downarrow
\rangle - \lvert \downarrow \uparrow \rangle) / \sqrt{2}$ and triplet,
$\lvert T_{0} \rangle := (\lvert \uparrow \downarrow \rangle + \lvert
\downarrow \uparrow \rangle) / \sqrt{2}$, states of the spin-zero
manifold. Designating $|\Scal_{0}\rangle = |S\rangle$ and
$|\Scal_{1}\rangle = |T_{0}\rangle $, we equate $C(t)$ in our general
model [eq.~\ref{eq:Hamiltonian}] with the exchange interaction.

The spin-zero manifold is insensitive to a global magnetic field.
However, gradients in the magnetic field will cause singlet-triplet
transitions. Such a gradient splits the energy of the states $\lvert
\uparrow \downarrow \rangle = (\lvert T_{0} \rangle + \lvert S \rangle)
/ \sqrt{2}$ and $\lvert \downarrow \uparrow \rangle = (\lvert T_{0}
\rangle - \lvert S \rangle) / \sqrt{2}$ by the difference in effective
Zeeman energies for an electron in either of the two quantum dots. In
this context, we may therefore equate the energy $\varepsilon$ with this
effective Zeeman energy difference. In GaAs DQD systems, the effective
Zeeman energy difference is typically dominated by the Overhauser shifts
from a lattice of randomly polarized nuclear spins corresponding to
approximately $1.602 \times 10^{-26}$ -- $1.602 \times 10^{-25}$ J (or
$10^{-7}$ -- $10^{-6}$ eV) \cite{Hanson07a, Taylor07a}. It has been
demonstrated that a desired difference in Overhauser shift of a GaAs DQD
may be realized through feedback control from a preparatory qubit
\cite{Bluhm10a}. However, the value of $\varepsilon$ will drift over
time through the nuclear spin diffusion that causes spectral diffusion
\cite{Witzel06a, Witzel07a, Yao06a, Liu07a}, motivating the need for
robust control. In proposed Si DQD systems, e.g., ref.~\cite{Levy09a},
nuclear spins may be eliminated through isotopic enrichment. Other spin
baths, such as electron spins of donor impurities \cite{Witzel10a} or
dangling bond spins at an interface \cite{deSousa07a}, may also lead to
variations and drift in the value of $\varepsilon$.

\subsection{Scaled unit system}
\label{ssec:units}

In addition to setting the reduced Planck constant $\hbar = 1$
(corresponding to unit of angular momentum: energy $\times$ time), a
simple set of scaled units is defined by also setting the final time of
the controlled evolution $\tf = 1$. The ratio $\hbar/\tf = 1$ yields the 
scaled unit of energy, which, when $\tf = 20$ ns (as an example),
corresponds to approximately $5.273 \times 10^{-27}$ J (or $3.291 \times
10^{-8}$ eV). By appropriately scaling the Hamiltonian and control
field, this propagation time can be transformed to any final time $\tf$.
Relationships between these scaled units and SI units (when $\tf = 20$
ns) are summarized in table \ref{tab:units}.

In this work, optimizations were performed for individual values of
$\varepsilon \in [0, 5]$; we denote the nominal values of
$\varepsilon$ used in these calculations as $\varepsilon_{0}$. This
range of $\varepsilon$ corresponds to no and moderate rotations from the
environment for the GaAS DQD example. For a DQD logical qubit,
\begin{equation}
\varepsilon = g_{\rme} \mu_{\rmB} B_{\! \Delta},
\end{equation}
where $g_{\rme}$ is the so-called electron $g$-factor, $\mu_{\rmB}$ is the
Bohr magneton, and $B_{\! \Delta}$ is the magnetic field resulting from
the difference in the random hyperfine fields from each quantum dot
along the direction of the applied field. When 1 scaled unit of time
corresponds to 20 ns (a representative estimate of the time required for
one-qubit rotations for a DQD system  \cite{Petta05a, Petta10a}),
$\varepsilon = 5$ scaled units of angular frequency (the maximum value
of $\varepsilon_{0}$ considered) corresponds to $B_{\! \Delta} \approx
6.5$ mT; this is consistent with experimental reports of GaAs DQDs
(where $g_{\rme} = -0.44$) \cite{Petta05a, Hanson07a, Taylor07a}. Unless
stated otherwise, all physical quantities in this work are expressed in
scaled units.

\begin{table}[ht]
\begin{tabular}{@{}|c|c|c|}
\hline
\ \ Physical quantity \ \ & \ \ Scaled unit \ \ & \ \ SI unit \ \ \\
\hline\hline
\ \ angular momentum: $\hbar$ \ \ & \ \ 1 \ \ & \ \ $1.055 \times
10^{-34}$ J\,s \ \ \\
\hline
time: $t$ & 1 & $2.0 \times 10^{-8}$ s \\
\hline
energy: $C$, $\varepsilon$ & 1 & $5.273 \times 10^{-27}$ J \\
\hline
\end{tabular}
\caption{Scaled and SI units for the logical qubit described by the
Hamiltonian in eq.~\eqref{eq:Hamiltonian}.} 
\label{tab:units}
\end{table}

\section{Optimal control of unitary operations via gradient-based
  algorithms}
\label{sec:control-GrA}

\subsection{Objective functionals for unitary operations}

For a target unitary operation $V \in \Urm(\Hcal)$, the distance
$\Delta$ between $V$ and a simulated final-time unitary operation
$U(\tf)$ is
\begin{subequations}
\label{eq:distance}
\begin{align}
\Delta[V, U(\tf; C)] & := \min_{\varphi \in \Rbb} \frac{1}{\sqrt{2n}} 
\left\| U(\tf; C) - \exp(\rmi \varphi) V \right\|_{\rHS}, \\
& = \sqrt{1 - \frac{1}{n} \left| \Tr \left[ V^{\dag} U(\tf) \right]
\right|},
\end{align}
\end{subequations}
where $\| \cdot \|_{\rHS}$ denotes the norm based on the Hilbert-Schmidt
inner product: $\langle A, B \rangle_{\rHS} := \Tr(A^{\dag} B)$, $A, B
\in \Mcal_{n}(\Cbb)$ [$\Mcal_{n} \left( \Xbb \right)$ denotes the set of
$n \times n$ matrices over the field $\Xbb$]. This
\emph{phase-invariant} distance measure is a special case of a more
general distance measure developed in ref.~\cite{Grace10a}, which is
applicable to studies involving composite systems where only the
qubit/system dynamics are directly of interest \cite{Grace07a,
Grace07b}.

Concerning mathematical notation, because the unitary time-evolution
operator is a function of time and a functional of the control, it will
be expressed more generally as $U(t; C)$ for all time $t$ and a control
$C$, compared to $U(t)$; the final-time unitary operator will be
expressed more generally as $\Utf(C)$, compared to $U(\tf)$. Also, we
denote the space of admissible controls with final time $t = \tf$ as
$\Cfrak_{\tf}$. Some properties of the Hilbert space $\Cfrak_{\tf}$ are
discussed below; further details are in ref.~\cite{Grace10a}.

Because $0 \leq \Delta \leq 1$ in general, it is useful to define the
fidelity $\Fcal$ of unitary operations as \cite{Fuchs99a, Grace10a}
\begin{equation}
\Fcal := \frac{1}{n} \left| \Tr \left[ V^{\dag} U(\tf) \right] \right|
= 1 - \Delta^{2}(V, \Utf),
\end{equation}
which is a common phase-invariant measure of gate fidelity based on the
Hilbert-Schmidt inner product, e.g., refs.~\cite{Palao03a, Dominy08a,
Ho09a}. Note the quadratic dependence of $\Fcal$ on $\Delta$, i.e., a
distance of $10^{x}$ corresponds to a fidelity of $1 - 10^{2x}$, where
$x \leq 0$.

An optimal control field for a given unitary operation may be located by
minimizing an objective functional $\Jcal[C]$ of the control field that
incorporates the final-time unitary target $V$, constrains the dynamics
of $U(t)$ to evolve according to eq.~\eqref{eq:Schrodinger}, and
penalizes the fluence of the control field. For this work, the objective
functional is defined as
\begin{equation}
\label{eq:objective}
\Jcal[C] := \Delta [V, \Utf(C)] + \frac{\alpha}{2} \int_{0}^{\tf} \!
\frac{C^{2}(t)}{s(t)} \rmd t.
\end{equation}
Often, the minimization of $\Jcal$ is performed using a gradient-based
algorithm (GrA; see \cite{Palao03a, Grace07a, Balint08a, Grace10a} for
details and examples of gradient-based optimizations). Here, $\alpha
\geq 0$ weighs the control-field fluence relative to the distance
$\Delta$ and $s : [0, \tf] \mapsto \Rbb$ is a continuous ``shape
function''. When appropriately chosen, $s(t)$ penalizes
undesirably-shaped functions \cite{Grace10a}. We use $s(t) =
\sin^{p}(\pi t/\tf)$, where $p \in \Qbb_{+} \cup \{0\}$. For $p \neq 0$,
this form penalizes the control-field slew rate around the initial and
final times and favors controls where $C(0) = C(\tf) = 0$.

For the time-dependent Hamiltonian in eq.~\eqref{eq:Hamiltonian}, $\Utf
: \Cfrak_{\tf} \to \Urm(\Hcal)$ denotes the map, defined implicitly
through the Schr\"{o}dinger equation [eq.~\eqref{eq:Schrodinger}], that
takes a control field $C \in \Cfrak_{\tf}$ to the final-time unitary
evolution operator $\Utf \in \Urm(\Hcal)$. Note that $\Cfrak_{\tf}$ is a
Hilbert space of admissible controls, on which $U(t; C)$ exists for all
$t \in [0, \tf]$ and all $C \in \Cfrak_{\tf}$ \cite{Jurdjevic72a}, where
the inner product on $\Cfrak_{\tf}$ is
\begin{equation}
\label{eq:control-space}
\langle f, g \rangle_{\Cfrak_{\tf}} := \int_{0}^{\tf} \!
\frac{f(t) g(t)}{s(t)} \rmd t, \ \ \forall f, g \in \Cfrak_{\tf}.
\end{equation}
As such, $\Jcal : \Cfrak_{\tf} \to \Rbb$ is the \emph{dynamical} version
of the distance measure $\Delta$, with a relative cost on the
control-field fluence, determined by $\alpha$. The role of the shape
function $s(t)$ in eqs.~\eqref{eq:objective} and
\eqref{eq:control-space} is to change the geometry of control space,
moving undesirably-shaped functions away from the origin, out to
infinity, where they are less likely to be the targets of a minimization
over $\Cfrak_{\tf}$ \cite{Grace10a}.

\subsection{\texorpdfstring{Control rotation angle $\bm{\theta(t; C)}$}
  {Control rotation angle}}
\label{ssec:control-theta}

In addition to the objective functional $\Jcal$ and inner product on
$\Cfrak_{\tf}$, another important expression is the integral of the
control field:
\begin{subequations}
\begin{equation}
\label{eq:theta}
\theta(t; C) := \int_{0}^{t} \! C(\tau) \rmd \tau.
\end{equation}
The angle $\theta$ corresponds to the rotation about the $z$-axis
performed by the control field during the time interval $[0, t]$.
Although $\theta$ is a functional of the control field $C(t)$, whenever
appropriate we abbreviate $\theta(t; C)$ as $\theta(t)$ to avoid
unnecessarily cumbersome notation. Equation~\eqref{eq:theta} is
equivalent to
\begin{equation}
\frac{\rmd \theta(t; C)}{\rmd t} = C(t).
\end{equation}
Also,
\begin{equation}
\frac{\delta \theta[t; C]}{\delta C(t')} = \Hfrak(t - t'),
\end{equation}
\end{subequations}
(i.e., the functional derivative of $\theta$ with respect to $C$) where
$\Hfrak(t - t')$ is the Heaviside step function:
\begin{equation}
\Hfrak(t' - t) :=
\left\{ \begin{array}{cc}
1 & \textrm{when} \ t \geq t' \\
0 & \textrm{when} \ t < t'
\end{array}\right. .
\end{equation}

\subsection{Optimization with a gradient-based algorithms}

This section briefly summarizes the variational analysis of $\Jcal$ and
describes criteria for the optimal points (or submanifolds) of $\Jcal$
with respect to a control field $C$. The gradient of the objective
functional $\Jcal$ is explicitly derived in ref.~\cite{Grace10a}; we
present it here for continuity:
\begin{equation}
\label{eq:gradient}
\big( \nabla \Jcal[C] \big)(t) = \frac{s(t)}{4n \, \Delta
[V, \Utf(C)]} \Imag \Big( \Tr \Big\{ \big[ \Utf^{\dag}(C) R - R^{\dag}
\Utf(C) \big] U^{\dag}(t; C) S_{z} U(t; C) \Big\} \Big) + \alpha C(t),
\end{equation}
where $R := \exp(\rmi \varphi) V$ and $\varphi := \Imag \left\{ \ln
\left[ \Tr \left( V^{\dag} \Utf \right) \right] \right\}$. Critical
points of $\Jcal$ (a real-valued functional) are defined as controls for
which $\big( \nabla \Jcal[C] \big)(t) = 0$ for all time $t$
\cite{Brif12a}. Control fields are iteratively updated using this
gradient to improve the value of the objective functional $\Jcal$. Given
the $k$th iterate of the control field $C^{(k)}(t)$, adjustments to the
control field for the $(k\!+\!1)$th iteration are given by
\begin{equation} 
\label{eq:field-adjust}
C^{(k+1)}(t) := C^{(k)}(t) - \beta \left( \nabla \Jcal \left[ C^{(k)}
\right] \right)(t),
\end{equation}
where $\beta$ is a constant that determines the magnitude of the field
adjustment. This procedure describes an implementation of a
steepest-descent algorithm \cite{Press92a}. In this work, initial
control fields $C^{(0)}$ are continuous approximations to simple
square-wave pulses, where initial and final times and slew rates are
consistent with the shape function $s(t) = \sin(\pi t/\tf)$.

\section{Results from quantum optimal control theory}
\label{sec:results-OCT}

Using \emph{only} the GrA presented in section \ref{sec:control-GrA},
OCs were found for unitary targets that perform $\pi/2$- and
$\pi$-rotations about the $z$-axis:
\begin{equation}
Z_{\phi} := \left( \begin{array}{cc} 
\exp \left( -\rmi \phi/2 \right) & 0 \\ 
0 & \exp \left( \rmi \phi/2 \right) 
\end{array} \right),
\end{equation}
where $\phi \in \{\pi/2, \pi\}$. The final time for all OCs was fixed
at $\tf = 1$ scaled unit of time. With the GrA and the objective
functional $\Jcal$, a combination of the value of $\varepsilon$ and the
structure of the initial control field determines the resulting optimal
control field. Optimizations were performed individually for
\emph{specific} angular frequencies: $\varepsilon_{0} \in [0, 5]$. As
described in section \ref{ssec:units}, this interval represents the
regime of zero to moderate rotation from the environment for the DQD
logical qubit. To emphasize the $\varepsilon$-specific nature of these
OCs, we denote them as $C_{\rmo}(\varepsilon_{0}; t)$.

\begin{figure}
\centering
\includegraphics[width=0.8\textwidth]{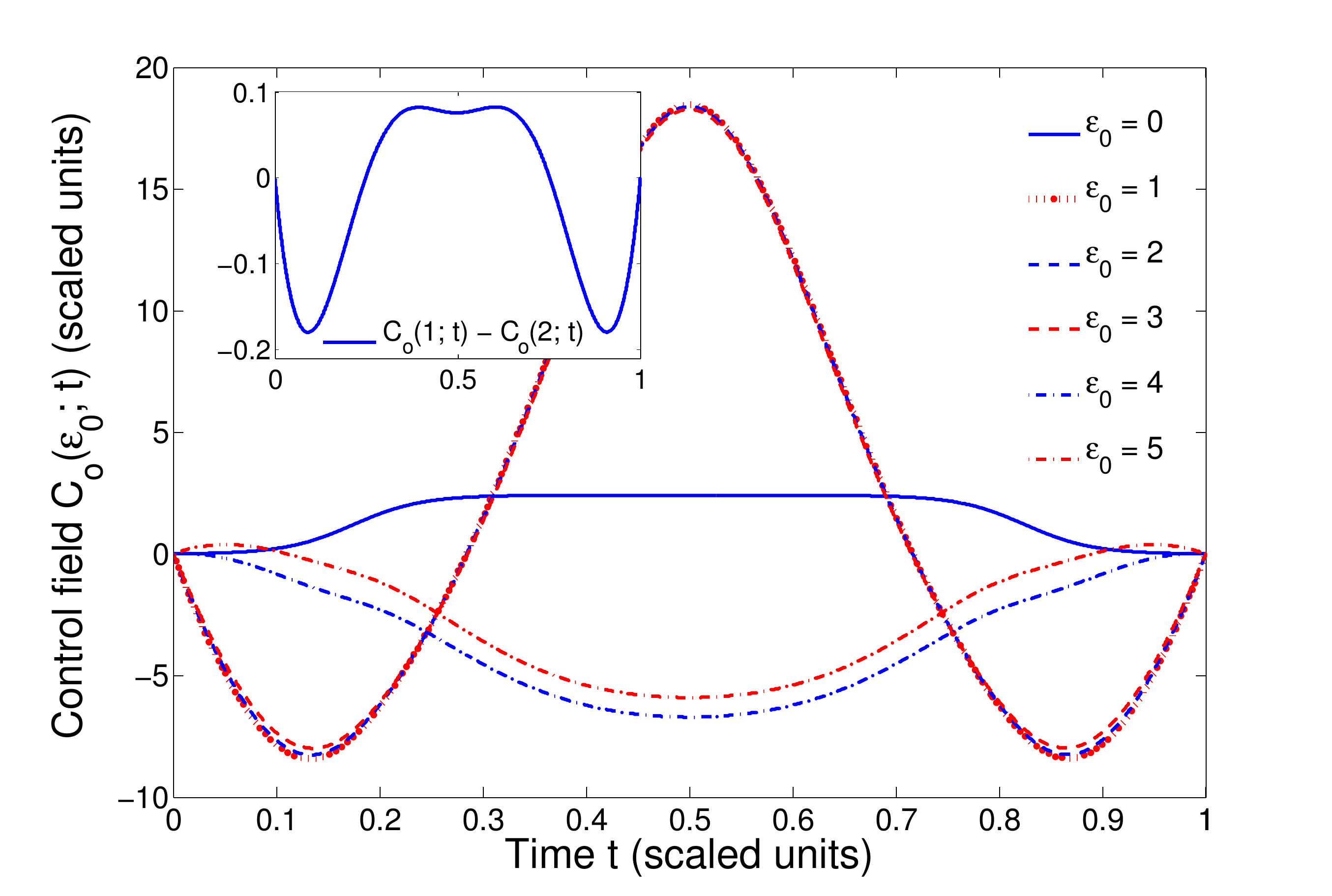}
\caption{(Color online) OC fields producing a $Z_{\pi/2}$ operation for
  several \emph{specific} values of $\varepsilon_{0}$ and $\tf = 1$
  scaled unit. All OCs have distances $\Delta(Z_{\pi/2}, \Utf) \leq
  10^{-6}$. The inset presents the difference between $C_{\rmo}(1; t)$
  and $C_{\rmo}(2; t)$. Although distinct, note that $C_{\rmo}(1; t)$,
  $C_{\rmo}(2; t)$, and $C_{\rmo}(3; t)$ appear nearly indistinguishable
  in this figure.}
\label{fig:fields-OCT-Z2}
\end{figure}

\begin{figure}
\centering
\includegraphics[width=0.8\textwidth]{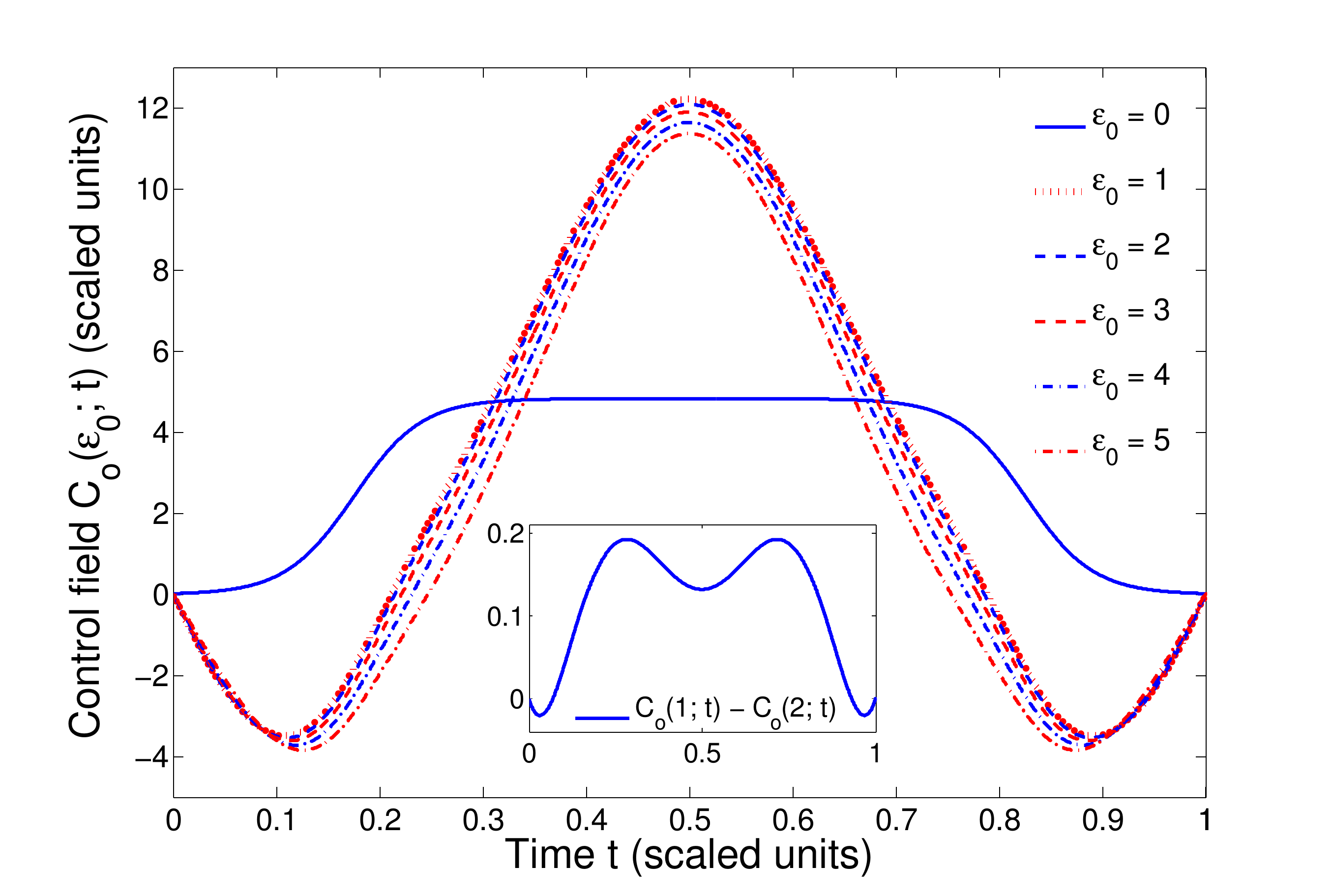}
\caption{(Color online) OC fields producing a $Z_{\pi}$ operation for
  several \emph{specific} values of $\varepsilon_{0}$ and $\tf = 1$
  scaled unit. All OCs have distances $\Delta(Z_{\pi}, \Utf) <
  10^{-6}$. The inset presents the difference between $C_{\rmo}(1; t)$
  and $C_{\rmo}(2; t)$.}
\label{fig:fields-OCT-Z}
\end{figure}

Because of the similarity of results over the entire interval $0 \leq
\varepsilon \leq 5$, only a subset will be presented. OC fields for
$Z_{\pi/2}$ and $Z_{\pi}$ as a function of $\varepsilon$ are presented in
figs.~\ref{fig:fields-OCT-Z2} and \ref{fig:fields-OCT-Z}, respectively,
for $\varepsilon_{0} \in \{ 0, 1, 2, 3, 4, 5 \}$. Even though all of
these OCs were located using the same initial control field, which is
very similar to the OC reported for $\varepsilon_{0} =  0$ for both
targets, some of the converged fields differ dramatically for different
values of $\varepsilon_{0}$. All OCs have distances $\Delta < 10^{-6}$
(or $\Fcal > 1 - 10^{-12}$, essentially corresponding to the limits of
numerical precision), which is expected because this system is
relatively simple and fully controllable \cite{Huang83a, Rama95a}. For
$\varepsilon_{0} = 0$, all OCs for $Z_{\phi}$ satisfy $\theta(\tf)
\equiv \phi \! \pmod{2\pi}$, i.e., OC design simply corresponds to
pulse-area control in this situation. However, when $\varepsilon \neq
0$, there is no corresponding pulse-area requirement. In fact, if
$\varepsilon$ is known accurately, it is possible to perform $Z_{\phi}$
operations with piecewise constant controls that satisfy $\theta(\tf) =
0$. Table \ref{tab:OCT-parameters} contains information about some of
the properties of these OCs. For a DQD logical qubit, we observe that
these controls require negative exchange coupling values. Although
negative exchange energy is uncommon, it is predicted to be possible to
produce through combined tuning of the magnetic field, dot size, and
tunnel coupling \cite{Nielsen10a}.

\begin{figure}
\centering
\includegraphics[width=0.8\textwidth]{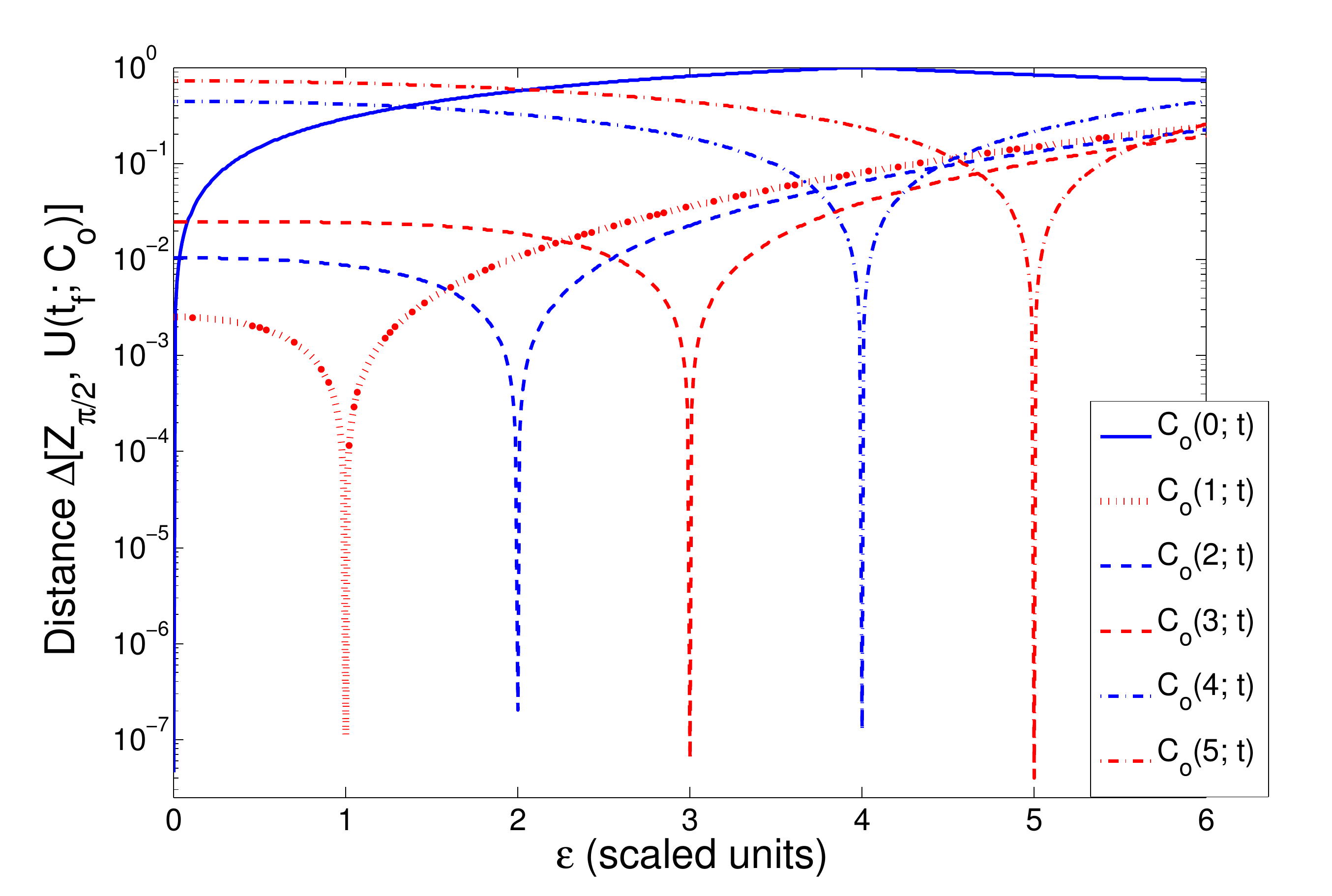}
\caption{(Color online) Distance of OCs optimized with \emph{particular}
  values of epsilon, denoted as $C_{\rmo}(\varepsilon_{0}; t)$, for the
  $Z_{\pi/2}$ operation [$\Delta(Z_{\pi/2}, \Utf) < 10^{-6}$ for all
  controls], subsequently applied over the interval $0 \leq \varepsilon
  \leq 6$ (with a resolution of 0.01 scaled units).}  
\label{fig:distance-OCT-Z2}
\end{figure}

\begin{figure}
\centering
\includegraphics[width=0.8\textwidth]{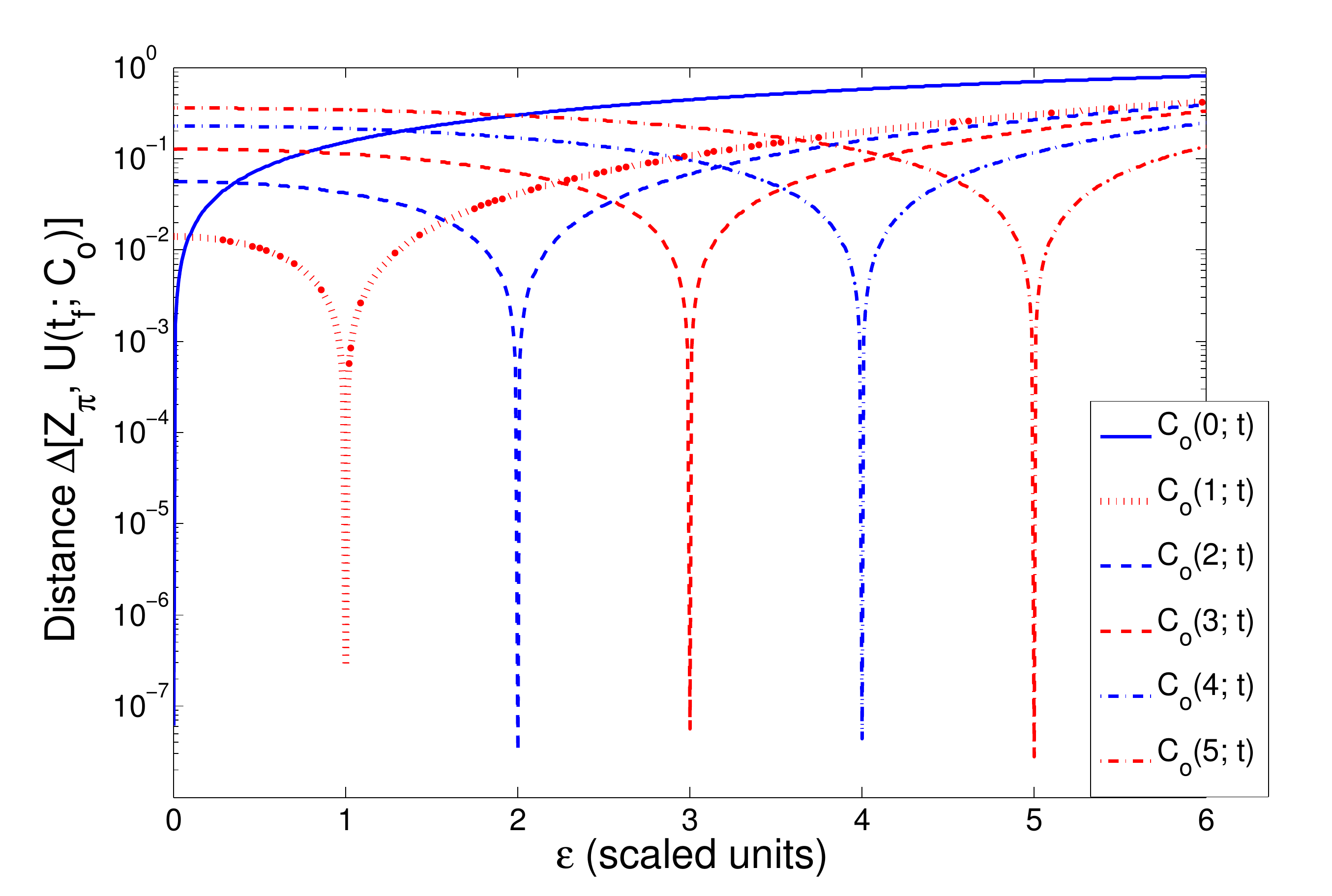}
\caption{(Color online) Distance of OCs optimized with \emph{particular}
  values of epsilon, denoted as $C_{\rmo}(\varepsilon_{0}; t)$, for the
  $Z_{\pi}$ operation [$\Delta(Z_{\pi}, \Utf) < 10^{-6}$ for all
  controls], subsequently applied over the interval $0 \leq \varepsilon 
  \leq 6$ (with a resolution of 0.01 scaled units).}
\label{fig:distance-OCT-Z}
\end{figure}

For both $Z_{\pi/2}$ and $Z_{\pi}$ operations, despite the similar
structures of the OCs, especially $C_{\rmo}(1; t)$ and $C_{\rmo}(2; t)$,
their gate-distance responses for $0 \leq \varepsilon \leq 6$ (with a
numerical resolution of 0.01 scaled units) are quite unique, as shown in
figs.~\ref{fig:distance-OCT-Z2} and \ref{fig:distance-OCT-Z}, 
respectively. Even though $\max_{t} \left| C_{\rmo}(1; t) - C_{\rmo}(2;
t) \right| < 0.2$ scaled units of energy for both operations (see inset
of figs.~\ref{fig:fields-OCT-Z2} and \ref{fig:fields-OCT-Z}), and the
mean relative difference is approximately 1.4\% and 5.7\% for the
$Z_{\pi/2}$ and $Z_{\pi}$ operations, respectively, this two-level
system effectively discriminates between these two similar control
fields through the response of the distance functional $\Delta$ in
eq.~\eqref{eq:distance}. Within the interval $0 \leq \varepsilon \leq
3$, the gate distances of $C_{\rmo}(1; t)$ and $C_{\rmo}(2; t)$ do not
significantly overlap. The sensitivity of this system to these
relatively small control-field variations combined with the inherent
noise (and limited resolution) present in most realistic control sources
warrants further study of the impact that realistic control-field
fluctuations may have on practical fault-tolerant QC \cite{Rabitz04a,
Dominy08a, Levy09a, Levy11a}.

For the target operation $Z_{\pi/2}$, when $\varepsilon_{0} \in \{ 0, 1,
2, 3 \}$, the OCs produce a net positive angle of rotation about the
$z$-axis, given the initial control field. When $\varepsilon_{0} \in \{
4, 5 \}$, OCs produce a net negative angle of rotation about the
$z$-axis. As the static angular frequency of the rotation about the
$x$-axis increases, the OC strategy tends toward a controlled rotation
about the $z$-axis in the negative direction. OC simulations for $3 \leq
\varepsilon_{0} \leq 4$, with a numerical resolution of 0.01 scaled
units of energy (detailed results are not reported), reveal a distinct
transition between OCs with shapes very similar to $C_{\rmo}(3, t)$ and
$C_{\rmo}(4, t)$ reported in fig.~\ref{fig:fields-OCT-Z2}, corresponding 
to net positive and negative rotations, respectively. For the initial
control field used in this work, this transition occurs when
$\varepsilon_{0} \approx 3.9$. Comparing the gate-distance responses in
fig.~\ref{fig:distance-OCT-Z2} for $C_{\rmo}(3, t)$ and $C_{\rmo}(4, t)$
reveals that these OCs are not equivalent solutions for unitary
control; the difference between these controls is exclusively due to the
effect of the different values of $\varepsilon_{0}$ on the qubit
dynamics.

Quantum-computing architectures often assume encoded quantum operations
to correct the inevitable errors due to control and environmental
noise. Gate operations, such as $Z_{\phi}$ simulated here, must
achieve a minimum fidelity threshold for successful quantum error
correction. A predicted maximum distance $\Delta$ of less than $10^{-3}$
for $\varepsilon \in [0, 5]$ is within typical ranges necessary for
fault-tolerant QC \cite{Taylor07a, Gaitan08a, Levy09a}. These results
highlight the potential advantage of using OCT if estimates of
Hamiltonian parameters are known well. However, these OCs are not
robust to uncertainty in the magnitude of $\varepsilon$; in fact, they
are highly sensitive to small perturbatives in $\varepsilon$.
Uncertainty can result from incomplete or poor system parameter
estimates as well as from dynamics of the environment \cite{Witzel06a,
Witzel10a}.

The objective functional $\Jcal$ in eq.~\eqref{eq:objective} does not
include criteria to evaluate control-field robustness with respect to
variations in $\varepsilon$. To investigate \emph{any} inherent
robustness quantitatively, OC fields optimized for particular values of
$\varepsilon$ [i.e., $C_{\rmo}(\varepsilon_{0}; t)$] were subsequently
applied to a surrounding interval of $\varepsilon_{0}$. Results are
presented in figs.~\ref{fig:distance-OCT-Z2} and
\ref{fig:distance-OCT-Z} for the $Z_{\pi/2}$ and $Z_{\pi}$ operations,
respectively. Consider the distance response of $C_{\rmo}(2; t)$ to
variations in $\varepsilon$, which varies substantially with respect to
variations that correspond to $\sim \!\! 1$ mT fluctuations
(corresponding to approximately $1.6 \leq \varepsilon \leq 2.4$) for
GaAS DQD systems. Without specifying a measure of robustness as an
additional control objective, the resulting OCs are not inherently
robust to modest local magnetic-field fluctuations (e.g., at each
quantum dot). Numerical calculations with these OCs for both operations
indicate that the error in the measurement fidelity used to characterize
$\varepsilon$ for a particular system must be smaller than $10^{-2}$
(corresponding to approximately $1.3 \times 10^{-5}$ or smaller T for
the GaAs DQD example) to realize gates with distances that are $10^{-3}$
or smaller. Moreover, with these controls, $\varepsilon$ could not drift
significantly during a computation without serious fidelity loss.

In addition to the data presented in figs.~\ref{fig:distance-OCT-Z2} and
\ref{fig:distance-OCT-Z}, we introduce the following functional to
investigate robustness of $Z_{\phi}$ operations over the interval
$[\varepsilon_{-}, \varepsilon_{+}]$:
\begin{equation}
\label{eq:robust-distance}
\Rcal_{\phi}[C, \varepsilon, \delta \varepsilon] :=
\int_{\varepsilon_{-}}^{\varepsilon_{+}} \!\!
\Delta \left[ Z_{\phi}, \Utf(C) \right] \rmd \varepsilon',
\end{equation}
where $\varepsilon_{\pm} := \varepsilon \pm \delta \varepsilon$. Values
for $\Rcal_{\phi}[C_{\rmo}, \varepsilon_{0}, 0.5]$, corresponding to the
average gate distance over a unit interval centered at
$\varepsilon_{0}$, are reported in table \ref{tab:OCT-parameters}.
Quantifying robustness with this metric further demonstrates the general
lack of robustness of these OCs; $\Rcal_{\phi}$ varies from $1.44 \times
10^{-3}$ (which is somewhat robust) to $7.40 \times 10^{-2}$.

\begin{table}
\begin{tabular}{@{}|c|c|c|c|c|c|c|}
\multicolumn{7}{c}{\textbf{Target operation:} $\bm{Z_{\pi/2}}$} \\
\hline
$\varepsilon_{0}$ & 0 & 1 & 2 & 3 & 4 & 5 \\
\hline\hline
\ $\max|C_{\rmo}|$ \ & 2.4 & 18.5 & 18.4 & 18.3 & 6.7 & 5.9 \\
\hline
$\theta(\tf; C_{\rmo})$ & $\pi/2$ & 1.5779 & 1.6002 & 1.6406 & -3.4333 &
-2.5660 \\
\hline
$\Phi[C_{\rmo}]$ & 3.4 & 84.5 & 83.0 & 80.7 & 17.5 & 12.1 \\
\hline
\ $\Delta(Z_{\pi/2}, \Utf)$ \ & \ $4.59 \times 10^{-8}$ \ &
\ $1.11 \times 10^{-7}$ \ & \ $2.03 \times 10^{-7}$ \ &
\ $6.23 \times 10^{-8}$ \ & \ $1.25 \times 10^{-7}$ \ &
\ $3.94 \times 10^{-8}$ \ \\
\hline
\ $\Rcal_{\pi/2}[C_{\rmo}, \varepsilon_{0}, 0.5]$ \ & \ $7.40 \times
10^{-2}$ \ &
\ $1.44 \times 10^{-3}$ \ & \ $3.65 \times 10^{-3}$ \ &
\ $6.96 \times 10^{-3}$ \ & \ $5.04 \times 10^{-2}$ \ & 
\ $6.21 \times 10^{-2}$ \ \\
\hline
\multicolumn{7}{c}{\textbf{Target operation:} $\bm{Z_{\pi}}$} \\
\hline
$\varepsilon_{0}$ & 0 & 1 & 2 & 3 & 4 & 5 \\
\hline\hline
\ $\max|C_{\rmo}|$ \ & 4.8 & 12.2 & 12.1 & 11.9 & 11.6 & 11.4 \\
\hline
$\theta(\tf; C_{\rmo})$ & $\pi$ & 3.1020 & 2.9824 & 2.7802 & 2.4923 &
2.1181 \\
\hline
$\Phi[C_{\rmo}]$ & 13.7 & 39.5 & 38.4 & 36.5 & 34.1 & 31.4 \\
\hline
\ $\Delta(Z_{\pi}, \Utf)$ \ & \ $6.05 \times 10^{-8}$ \ &
\ $2.73 \times 10^{-7}$ \ & \ $3.16 \times 10^{-8}$ \ &
\ $5.58 \times 10^{-8}$ \ & \ $4.34 \times 10^{-8}$ \ &
\ $2.79 \times 10^{-8}$ \ \\
\hline
\ $\Rcal_{\pi}[C_{\rmo}, \varepsilon_{0}, 0.5]$ \ & \ $3.80 \times
10^{-2}$ \ &
\ $6.95 \times 10^{-3}$ \ & \ $1.38 \times 10^{-2}$ \ &
\ $2.05 \times 10^{-2}$ \ & \ $2.68 \times 10^{-2}$ \ & 
\ $3.24 \times 10^{-2}$ \ \\
\hline
\end{tabular}
\caption{Performance of the OCs $C_{\rmo}(\varepsilon_{0}; t)$ for
  one-qubit $Z_{\phi}$ operations. Here, $\max|C_{\rmo}|$, $\theta$,
  $\Phi[C] := \int_{0}^{\tf} \! C^{2}(\varepsilon_{0}; t) \rmd t$,
  $\Delta$, and $\Rcal_{\phi}$ are the maximum control-field amplitude, angle
  of controlled $z$-axis rotation, control-field fluence, gate distance,
  and gate robustness, respectively, in the corresponding scaled units
  described in section \ref{ssec:units}.}
\label{tab:OCT-parameters}
\end{table}

\section{Robust decoupling-pulse criteria}
\label{sec:robust-DPC}

Optimization of the functional $\Jcal$ in eq.~\eqref{eq:objective} is
highly under-determined, and multiple control fields exist that will
produce the same target operator $V$ \cite{Ho09a, Dominy11a}. Requiring
robustness to control and/or system variations involves the
specification of additional constraints or penalties, such as
eq.~\eqref{eq:robust-distance}, thereby limiting solutions to this OCT
problem. In this section, we summarize a set of control-field
constraints that characterize robustness to perturbative decoherence and
adapt them to locate controls that are robust to system uncertainty.

\subsection{General robustness criteria}

Consider the following open-system Hamiltonian for one qubit:
\begin{equation}
\label{eq:Hamiltonian-open1}
H_{\mathrm{open}}(t) := \vec{S} \cdot \vec{C}(t) + \vec{S} \cdot
\vec{\Gamma} + H_{\rme},
\end{equation}
where $\vec{S} := (S_{x}, S_{y}, S_{z})$ represents the spin-operator
vector, $\vec{C}(t) := (C_{x}, C_{y}, C_{z})$ represents a multi-polarized
control field, $\vec{\Gamma} := (\Gamma_{x}, \Gamma_{y}, \Gamma_{z})$
represents the environment interaction operator, and $H_{\rme}$
represents the environment Hamiltonian.

By expanding the final-time unitary evolution operator generated by
$H_{\mathrm{open}}$ with respect to $\tf \| H_{\rme} \|$ and $\tf \|
\vec{\Gamma} \|$ about $\tf \| H_{\rme} \| = 0$ and $\tf \| \vec{\Gamma}
\| = 0$, Pasini \emph{et al.}~have identified control-field criteria
necessary to eliminate perturbative first- and second-order effects
resulting from the environment Hamiltonian $H_{\rme}$ and the
interaction term $\vec{S} \cdot \vec{\Gamma}$ \cite{Pasini09a}. Although
applicable to multi-polarized controls and general qubit-environment
coupling, the methodology developed in this section assumes
control-qubit coupling along the $z$-axis and qubit-environment
interaction along the $x$-axis, i.e.,
\begin{equation}
\label{eq:Hamiltonian-open2}
H_{\mathrm{open}}'(t) := C(t) S_{z} \otimes \openone_{n_{\rme}} + S_{x} \otimes
\Gamma_{x} + \openone_{2} \otimes H_{\rme},
\end{equation}
where $n_{\rme} := \dim \{ \Hcal_{\rme} \}$ and $\Hcal_{\rme}$ is the
Hilbert space of the environment. For controlled $\pi/2$- and $\pi$
rotations about the $z$-axis, Pasini \emph{et al.}~derived the following
vector functional characterizing the space of controls that suppress
first- and second-order effects errors resulting from an $x$-axis
interaction with the environment:
\begin{subequations}
\label{eq:zeta}
\begin{gather}
\vec{\zeta}[\theta] := (\zeta_{1}, \zeta_{2}, \zeta_{3}, \zeta_{4},
\zeta_{5})^{\rmT},
\intertext{where}
\zeta_{1}[\theta] := \int_{0}^{\tf} \! \sin[\theta(t)] \rmd t, \\
\zeta_{2}[\theta] := \int_{0}^{\tf} \! \cos[\theta(t)] \rmd t, \\
\zeta_{3}[\theta] := \int_{0}^{\tf} \!\! \int_{0}^{\tf} \!
\sin[\theta(t_{1}) - \theta(t_{2})] \sign(t_{1} - t_{2}) \rmd t_{1} \rmd
t_{2}, \\
\zeta_{4}[\theta] := \int_{0}^{\tf} \! t \sin[\theta(t)] \rmd t, \\
\zeta_{5}[\theta] := \int_{0}^{\tf} \! t \cos[\theta(t)] \rmd t.
\end{gather}
\end{subequations}
Recall $\theta(t; C)$ from eq.~\eqref{eq:theta} in
section \ref{ssec:control-theta}, which represents the net rotation
performed by the control field $C$ during the time interval $[0, t]$.

For convenience and simplicity in the analysis that follows this
section, we define $\vec{\eta}$ as
\begin{subequations}
\label{eq:eta}
\begin{gather}
\vec{\eta}[C] := (\eta_{1}, \eta_{2}, \eta_{3}, \eta_{4},
\eta_{5})^{\rmT},
\intertext{where}
\eta_{i} := \zeta_{i} \circ \theta.
\end{gather}
\end{subequations}
Thus, $\vec{\eta} : \Cfrak_{\tf} \mapsto \Rbb^{5}$. For one qubit, the
components of $\vec{\eta}$ represent the first- and second-order
perturbative errors, with respect to the final time $\tf$ and error
Hamiltonians $\Gamma_{x}$ and $H_{\rme}$, of a controlled $\pi/2$- or
$\pi$-rotation about the $z$-axis. Specifically, $\eta_{1}$ and
$\eta_{2}$ represent first-order errors, while $\eta_{3}$, $\eta_{4}$,
and $\eta_{5}$ represent second-order errors. Thus, when $\vec{\eta} =
0$, the pulse is accurate up to third-order, eliminating the first- and
second-order effects resulting from a perturbative qubit-environment
interaction. According to the analysis in ref.~\cite{Pasini09a}, when
$[H_{\rme}, \Gamma_{\lambda}] = 0$, for all $\lambda$, components
$\eta_{4}$ and $\eta_{5}$ can be neglected from the vector constraint.

\subsection{Closed-system robustness criteria}

To apply these results to a closed one-qubit system and construct robust
$Z_{\phi}$ operations for the DQD logical qubit using this criteria,
we first compare the Hamiltonians $H$ in eq.~\eqref{eq:Hamiltonian} and
$H_{\mathrm{open}}'$ in eq.~\eqref{eq:Hamiltonian-open2}. These
Hamiltonians are equal if $\Gamma_{x} = \varepsilon$ and $H_{\rme} = 0$,
which implies that $[H_{\rme}, \Gamma_{\lambda}] = 0$, so
$\vec{\eta}^{\, \rmr} := (\eta_{1}, \eta_{2}, \eta_{3})^{\rmT}$ is the
relevant \emph{reduced} vector constraint. Incorporating these nonlinear
equality constraints into the original optimization problem yields the
following nonlinearly-constrained control problem:
\begin{gather}
\nonumber \min_{C \in \Cfrak_{\tf}} \Jcal[C] \\
\label{eq:constrained-optimization}
\textrm{subject to} \\
\nonumber \vec{\eta}^{\, \rmr}[C] = 0
\end{gather}
Methods such as ``diffeomorphic modulation under
observable-response-preserving homotopy'' (DMORPH) \cite{Rothman05a,
Rothman05b, Rothman06a, Dominy08a} or sequential quadratic
programming \cite{Boggs95a} are required to generate OCs that maintain
or satisfy approximate feasibility, determined by $\vec{\eta}^{\, \rmr}
= 0$. A technique using DMORPH is developed and applied in the next
sections.

\begin{figure}
\centering \includegraphics[width=0.8\textwidth]{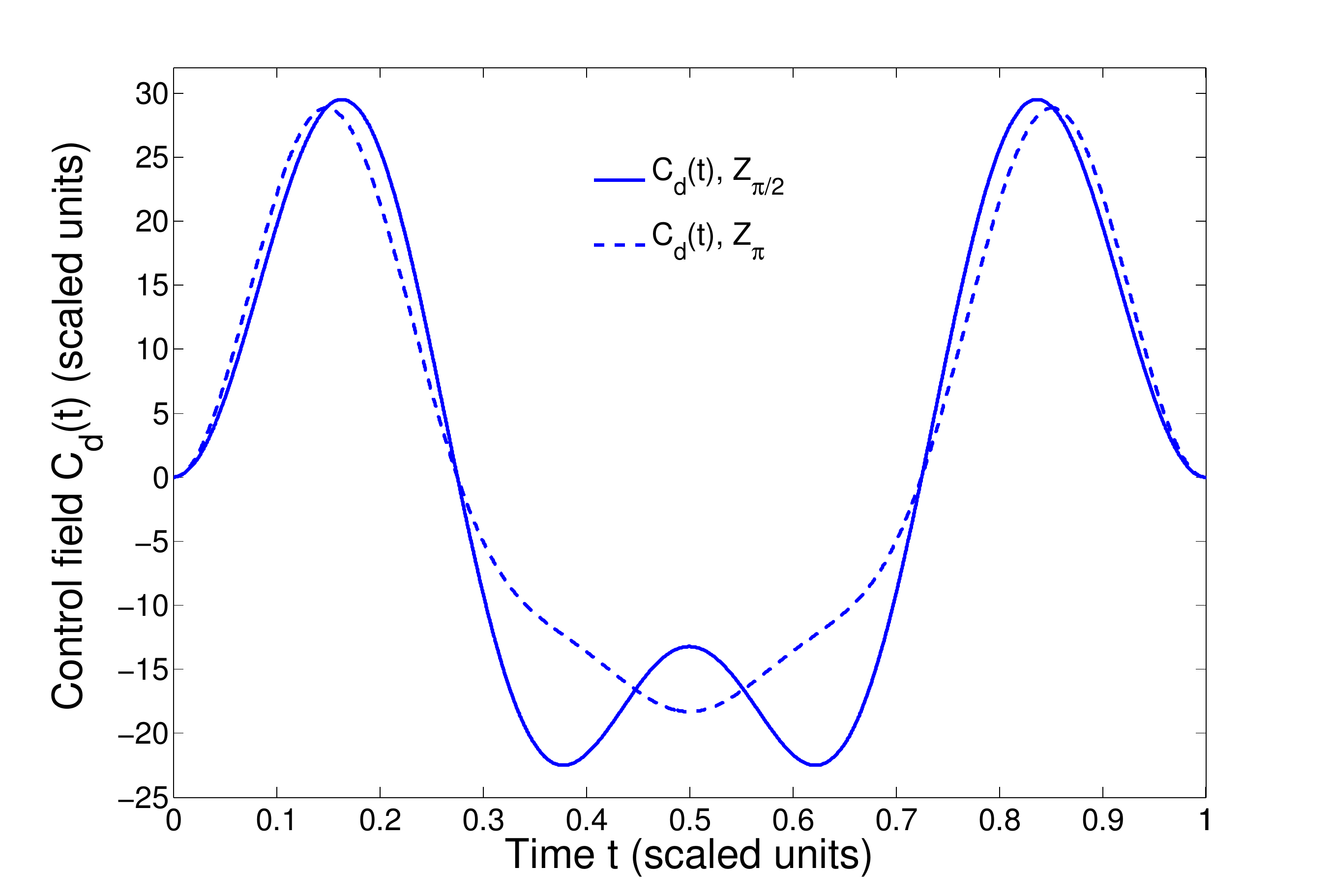}
\caption{(Color online) Control fields satisfying the DPC in
  eq.~\eqref{eq:eta}, denoted as $C_{\rmd}(t)$, for $Z_{\phi}$ operations
  ($\vec{\eta}^{\, \rmr} \|_{2}  < 10^{-7}$), from Pasini \emph{et
  al.}~\cite{Pasini09a}.}
\label{fig:fields-DPC-Z2-Z}
\end{figure}

\begin{figure}
\centering
\includegraphics[width=0.8\textwidth]{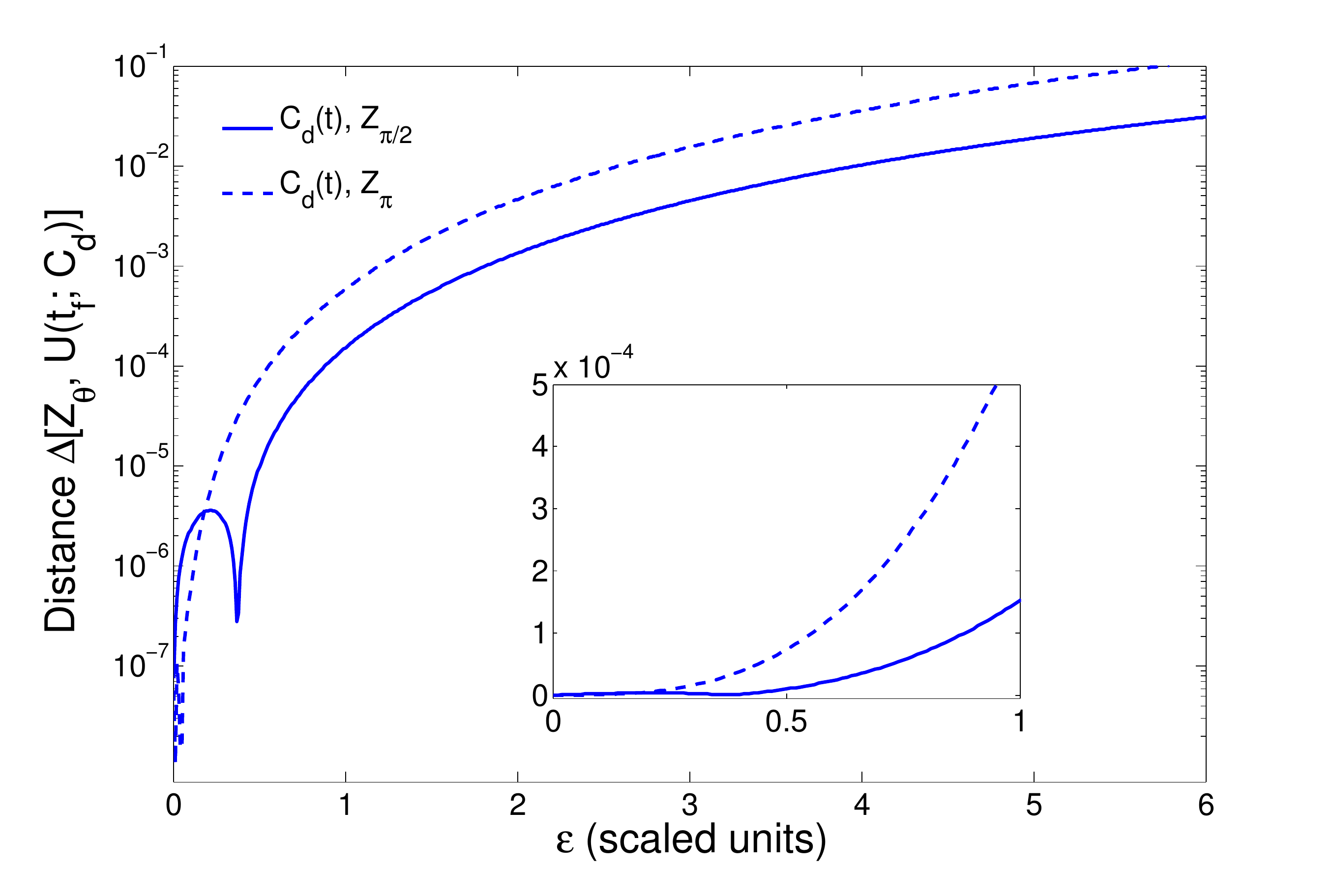}
\caption{(Color online) Distance of the $Z_{\phi}$ operations as a
  function of $\varepsilon$  (with a resolution of 0.01 scaled units)
  and $\phi \in \{\pi/2, \pi\}$ for the Landau-Zener model of
  eq.~\eqref{eq:Hamiltonian} and control fields $C_{\rmd}(t)$ presented
  in fig.~\ref{fig:fields-DPC-Z2-Z}. The inset displays the gate
  distance for $0 \leq \varepsilon \leq 1$ in greater detail, on a
  linear scale.}
\label{fig:distance-DPC-Z2-Z}
\end{figure}

For a qubit described by the Hamiltonian in eq.~\eqref{eq:Hamiltonian},
control fields from ref.~\cite{Pasini09a} satisfying $\| \vec{\eta}^{\,
\rmr} \|_{2} < 10^{-7}$ (where $\| \cdot \|_{2}$ denotes the vector
two-norm) and corresponding gate-distances for $\pi/2$- and $\pi$-pulses
are presented in figs.~\ref{fig:fields-DPC-Z2-Z} and
\ref{fig:distance-DPC-Z2-Z}, respectively. These fields are denoted as
$C_{\rmd}(t)$, where the subscript ``d'' indicates the decoupling
feature of these DPs. Satisfying the control-field constraints specified
by Pasini \emph{et al.}~\cite{Pasini09a}, first- and second-order
perturbations about $\varepsilon = 0$ are eliminated. As such, gate
distance increases with the magnitude of $\varepsilon$ and optimum
performance occurs when $\varepsilon = 0$, which is not necessarily
expected to be valid for realistic qubit systems with drift terms, e.g.,
\cite{Petta05a}. Although we use $\vec{\eta}^{\, \rmr} = 0$ as a general
condition for robustness in this work, whether controls satisfying
$\vec{\eta}^{\, \rmr} = 0$ are robust abouts points where $\varepsilon
\neq 0$ remains an open question. However, comparing the gate distances
in fig.~\ref{fig:distance-DPC-Z2-Z} to figs.~\ref{fig:distance-OCT-Z2}
and \ref{fig:distance-OCT-Z} reveals a certain degree of robustness in
control fields $C_{\rmd}(t)$ relative to $C_{\rmo}(\varepsilon; t)$,
e.g., for both $Z_{\phi}$ operations and all values of $\varepsilon_{0}$
considered, $\rmd \Delta / \rmd \varepsilon$ around $\varepsilon_{0}$
for $C_{\rmd}(t)$ is much smaller than $C_{\rmo}(\varepsilon_{0};
t)$. Table \ref{tab:DPC-parameters} contains information about some of
the properties of the DPs in figs.~\ref{fig:fields-DPC-Z2-Z} and
\ref{fig:distance-DPC-Z2-Z}.

\begin{table}
\begin{tabular}{@{}|c|c|c|c|c|c|c|}
\multicolumn{7}{c}{\textbf{Target operation:} $\bm{Z_{\pi/2}}$} \\
\hline
$\varepsilon_{0}$ & 0 & 1 & 2 & 3 & 4 & 5 \\
\hline\hline
\ $\Delta(Z_{\pi/2}, \Utf)$ \ & \ $7.60 \times 10^{-8}$ \ &
\ $1.52 \times 10^{-4}$ \ & \ $1.34 \times 10^{-3}$ \ &
\ $4.47 \times 10^{-3}$ \ & \ $1.02 \times 10^{-2}$ \ &
\ $1.89 \times 10^{-2}$ \ \\
\hline
\ $\Rcal_{\pi/2}[C_{\rmd}, \varepsilon_{0}, 0.5]$ \ & \ $2.84 \times
10^{-6}$ \ &
\ $1.96 \times 10^{-4}$ \ & \ $1.42 \times 10^{-3}$ \ &
\ $4.58 \times 10^{-3}$ \ & \ $1.03 \times 10^{-2}$ \ & 
\ $1.91 \times 10^{-2}$ \ \\
\hline
\end{tabular}
\begin{tabular}{@{}|c|c|c|c|}
\hline
\ \ $\max|C_{\rmd}| = 29.5$ \ \ & \ \ $\theta(\tf; C_{\rmd}) = \pi/2$
\ \ & \ \ $\Phi[C_{\rmd}] = 335.5$ \ \ & \ \
$\| \vec{\eta}^{\, \rmr}(t; C_{\rmd}) \|_{2} = 2.52 \times 10^{-8}$ \ \ \\
\hline
\end{tabular}
\begin{tabular}{@{}|c|c|c|c|c|c|c|}
\multicolumn{7}{c}{\textbf{Target operation:} $\bm{Z_{\pi}}$} \\
\hline
$\varepsilon_{0}$ & 0 & 1 & 2 & 3 & 4 & 5 \\
\hline\hline
\ $\Delta(Z_{\pi}, \Utf)$ \ & \ $5.67 \times 10^{-8}$ \ &
\ $5.84 \times 10^{-4}$ \ & \ $4.63 \times 10^{-3}$ \ &
\ $1.54 \times 10^{-2}$ \ & \ $3.56 \times 10^{-2}$ \ &
\ $6.75 \times 10^{-2}$ \ \\
\hline
\ $\Rcal_{\pi}[C_{\rmd}, \varepsilon_{0}, 0.5]$ \ & \ $1.83 \times
10^{-5}$ \ &
\ $7.29 \times 10^{-4}$ \ & \ $4.91 \times 10^{-3}$ \ &
\ $1.58 \times 10^{-2}$ \ & \ $3.61 \times 10^{-2}$ \ & 
\ $6.81 \times 10^{-2}$ \ \\
\hline
\end{tabular}
\begin{tabular}{@{}|c|c|c|c|}
\hline
\ \ $\max|C_{\rmd}| = 28.8$ \ \ & \ \ $\theta(\tf; C_{\rmd}) = \pi$ \ \ &
\ \ $\Phi[C_{\rmd}] = 264.8$ \ \ & \ \ 
$\| \vec{\eta}^{\, \rmr}(t; C_{\rmd}) \|_{2} = 8.04 \times 10^{-8}$ \ \ \\
\hline
\end{tabular}
\caption{Performance of the DPs $C_{\rmd}(t)$ for one-qubit $Z_{\phi}$
  operations. Here, $\Delta$, $\Rcal_{\phi}$, $\max|C_{\rmd}|$, $\theta$,
  $\Phi[C] := \int_{0}^{\tf} \! C^{2}(\varepsilon_{0}; t) \rmd t$, and
  $\| \vec{\eta}^{\, \rmr} \|_{2}$ are the gate distance, gate
  robustness, maximum control-field amplitude, angle of controlled
  $z$-axis rotation, control-field fluence, and constraint vector norm,
  respectively, in the corresponding scaled units described in
  section \ref{ssec:units}.}
\label{tab:DPC-parameters}
\end{table}

\section{Hybrid quantum control: decoupling-pulse criteria + optimal
  control theory}
\label{sec:DPC+OCT}

Given the favorable structure of quantum-control landscapes, e.g.,
trap-free structure, continua corresponding to optimal solutions,
etc., for regular controls \cite{Rabitz04a, Ho09a, Dominy11a, Moore11b},
DMORPH provides a mathematical means to explore families of controls
that achieve the same objective \cite{Rothman05a, Dominy08a}.
Applications of DMORPH include the continuous variation of a Hamiltonian
while preserving or optimizing the value of a quantum-mechanical
observable \cite{Rothman05a, Donovan11a} and exploring the level sets of
state and unitary control \cite{Rothman06a, Beltrani07a, Dominy08a}.
DMORPH can also be used as direct optimization technique
\cite{Moore11a}. We develop DMORPH techniques to explore the space of 
controls corresponding to $\vec{\eta} = 0$ while optimizing $\Jcal$ for
a specified $\varepsilon_{0}$.

Expressed more formally, in this section, we develop a method to
optimize $\Jcal$ over the set $\Cfrak_{\tf}^{\eta} := \vec{\eta}^{\,
-1}(0) \subset \Cfrak_{\tf}$, i.e., the set of feasible controls
satisfying $\vec{\eta} = 0$, where $\vec{\eta}^{-1}$ denotes the
preimage of $\vec{\eta}$ \footnote{The preimage of a particular subset
$S \subset Y$ of the codomain of a function $f: X \mapsto Y$ is the set
of all elements of the domain $X$ of $f$ that map to elements of $S$,
i.e., $f^{-1}(S) := \{x \in X : f(x) \in S\}$. Because $\vec{\eta} :
\Cfrak_{\tf} \mapsto \Rbb^{5}$, this implies that $\vec{\eta}^{-1} :
\Rbb^{5} \mapsto \Cfrak_{\tf}$.}. To increase general applicability, we
develop this technique for $\vec{\eta}$, rather than the reduced
constraint $\vec{\eta}^{\, \rmr}$. Away from critical points of
$\vec{\eta}$ (a real-valued vector), i.e., controls for which the set of
gradients $\{ \nabla \eta_{i} \}$ are linearly dependent
\cite{Guillemin74a}, $\Cfrak_{\tf}^{\eta}$ is a codimension 5
submanifold of $\Cfrak_{\tf}$ \cite{Abraham88a}. It is assumed that 
critical points of $\vec{\eta}$ are rare, an assumption supported by the
success of the resulting algorithm. The gradient of the restricted
functional $\Kcal := \Jcal\big|_{\Cfrak_{\tf}^{\eta}}$ at a point $C \in
\Cfrak_{\tf}^{\eta}$ is just the projection of the gradient of $\Jcal$
at $C$ onto the tangent space $\rmT_{C}\Cfrak_{\tf}^{\eta}$ of
$\Cfrak_{\tf}^{\eta}$ at $C$ \cite{doCarmo92a, Milnor97a, Dominy08a}. By
systematically updating the control field iteratively using a GrA with
this projected gradient, the algorithm is able to simultaneously improve
the value of $\Jcal$ and maintain approximate feasibility, or at least
impede deviations from feasibility. It is unlikely that the
quantum-control landscape for the restricted objective functional
$\Kcal$ is trap-free. As such, a global optimization algorithm
\emph{might} be better suited to finding solutions. However, because a
global parameterization of the set $\Cfrak_{\tf}^{\eta}$ is lacking,
maintaining (approximate) feasibility, i.e., $\vec{\eta} = 0$, might be
difficult in general.

\subsection{Gradients of the feasibility constraints}

Using DMORPH to remove the components of $\nabla \Jcal$ that cause a
change in $\vec{\eta}$ requires the gradient of each element of
$\vec{\eta}$, $\nabla \eta_{i}$:
\begin{subequations}
\label{eq:grad-eta}
\begin{gather}
\left( \nabla \eta_{i}[C] \right)(t) = \int_{0}^{\tf} \!
\frac{\delta \zeta_{i}[\theta]}{\delta \theta(\tau)} \times \frac{\delta
\theta(\tau)}{\delta C(t)} \rmd \tau = \int_{t}^{\tf} \!
\frac{\delta \zeta_{i}[\theta]}{\delta \theta(\tau)} \rmd \tau \ 
\Rightarrow \\ 
\left( \begin{array}{c}
\displaystyle{\left( \nabla \eta_{1}[C] \right)(t)} \\
\vdots \\
\displaystyle{\left( \nabla \eta_{5}[C] \right)(t)}
\end{array} \right) = 
\left( \begin{array}{c}
\displaystyle{\int_{t}^{\tf} \! \cos[\theta(\tau)] \rmd \tau} \\
\displaystyle{\int_{t}^{\tf} \! \sin[\theta(\tau)] \rmd \tau} \\
\displaystyle{2 \int_{t}^{\tf} \!\! \int_{0}^{\tf} \!\! \big\{
\cos[\theta(\tau)] \cos[\theta(\tau')] + \sin[\theta(\tau)]
\sin[\theta(\tau')] \big\} \sign(\tau - \tau') \rmd \tau' \rmd \tau} \\
\displaystyle{\int_{t}^{\tf} \! \tau \cos[\theta(\tau)] \rmd \tau} \\
\displaystyle{\int_{t}^{\tf} \! \tau \sin[\theta(\tau)] \rmd \tau}
\end{array} \right).
\end{gather}
\end{subequations}
As expressed here, the vector of gradients $\nabla \eta_{i}$ is a
function of the time variable $t$. Note that the set $\{ \nabla \eta_{i}
\}$ spans the normal space $\left( \rmT_{C} \Cfrak_{\tf}^{\eta}
\right)^{\perp}$ when $C \in \Cfrak_{\tf}^{\eta}$ is a regular point of
$\vec{\eta}$.

\subsection{Gradient projection method}

In addition to the gradients $\nabla \eta_{i}$, we also need a vector
that specifies the relative weight of each gradient component to
remove. This is determined by first calculating the Gramian matrix, with
elements
\begin{equation}
\label{eq:Gramian}
\left( G_{c} \right)_{ij} := \left\langle \nabla \eta_{i}[C], \nabla
\eta_{j}[C] \right\rangle_{\Cfrak_{\tf}}.
\end{equation}
In general, $G_{c}$ is not guaranteed to be full rank; non-singularity
of $G_{c}$ must be explored (numerically) as a function of $C$. The
Gramian matrix $G_{c}$ is rank deficient if and only if elements in the
set $\{ \nabla \eta_{i} \}$ are linearly dependent, i.e., if and only if
$C$ is a critical point of $\vec{\eta}$ \cite{Guillemin74a}. However,
when $\nabla \eta_{i}$ are all linearly independent, $G_{c}$ is
invertible. With eqs.~\eqref{eq:grad-eta} and \eqref{eq:Gramian}, all
gradient directions $\nabla \eta_{i}$ can be removed from $\nabla
\Jcal$, producing $\nabla \Kcal$ as follows:
\begin{subequations}
\label{eq:project}
\begin{equation}
\nabla \Kcal[C] = \nabla \Jcal[C] - \sum_{i = 1}^{5} \nabla \eta_{i}[C]
\left\{ G_{c}^{-1} \left[ \vec{q}_{c} \left( \nabla \Jcal[C] \right)
\right] \right\}_{i},
\end{equation}
where $\Kcal : \Cfrak_{\tf}^{\eta} \to \Rbb$ is a restriction of
$\Jcal$, i.e., $\Kcal := \Jcal|_{\Cfrak_{\tf}^{\eta}}$, and
$\vec{q}_{c}$ has elements
\begin{equation}
\left[ \vec{q}_{c} \left( \nabla \Jcal[C] \right) \right]_{i} :=
\left\langle \nabla \Jcal[C], \nabla \eta_{i}[C]
\right\rangle_{\Cfrak_{\tf}}.
\end{equation}
\end{subequations}

Thus, $\nabla \Kcal$ is a vector field on $\Cfrak_{\tf}^{\eta}$, and the
ordinary differential equation (ODE)
\begin{equation}
\label{eq:DMORPH-gradflow}
\frac{\rmd C(s)}{\rmd s} = -\nabla \Kcal[C(s)]
\end{equation}
describes the gradient flow of $\Kcal$ on $\Cfrak_{\tf}^{\eta}$ that
minimizes $\Jcal$ without changing the value of $\vec{\eta}$. The GrA in
this work implements a forward Euler integration of this equation, which
should be sufficiently accurate, provided that the multiplier $\beta$ in
eq.~\eqref{eq:field-adjust} is selected properly, i.e., $\beta$ is 
within the validity of the linear approximation of the tangent space at
$C$. Higher-order numerical ODE solvers, e.g., Runge-Kutta methods
\cite{Press92a}, might offer higher accuracy and/or greater efficiency,
but have not been explored in this work.

Equation \eqref{eq:project} describes the orthogonal projection from
$\rmT_{C} \Cfrak_{\tf}$ to $\rmT_{C} \Cfrak_{\tf}^{\eta}$. That
$\nabla \Kcal$ is orthogonal to all elements of $\{ \nabla \eta_{i} \}$
can be verified as follows. Let
\begin{equation}
\xi := \sum_{i = 1}^{5} \chi_{i} \nabla \eta_{i}[C],
\end{equation}
i.e., $\xi$ is a linear combination of the elements of the set $\left\{
\nabla \eta_{i} \right\}$, where $\chi_{i} \in \Rbb$. Replacing $\nabla
\Jcal$ with $\xi$ in eq.~\eqref{eq:project} yields
\begin{equation}
\xi - \sum_{i = 1}^{5} \nabla \eta_{i}[C] \left\{ G_{c}^{-1} \left[
\vec{q}_{c} \left( \xi \right) \right] \right\}_{i} = \xi - \sum_{i =
1}^{5} \nabla \eta_{i}[C] \left(G_{c}^{-1} G_{c} \vec{\chi} \right)_{i}
= 0,
\end{equation}
where $\vec{\chi} := (\chi_{1}, \chi_{2}, \chi_{3}, \chi_{4},
\chi_{5})^{\rmT}$. If $\xi \in \rmT_{C}\Cfrak_{\tf}$ is such that $\xi$
is orthogonal to all gradients $\nabla \eta_{i}$, then the projection
described in eq.~\eqref{eq:project} acts as identity on $\xi$. Together
with the previous statement, this shows that eq.~\eqref{eq:project} is
the orthogonal projector from $\rmT_{C} \Cfrak_{\tf}$ to $\rmT_{C}
\Cfrak_{\tf}^{\eta}$.

As mentioned in the introduction, because we combine DPC and OCT
methods to generate improved control fields, we denote the integrated
optimization procedure described in this section as DPC+OCT.
Straightforward modifications of $G_{c}$ and $\vec{q}_{c}$ are required 
when $\vec{\eta}^{\, \rmr}$ is the constraint vector rather than
$\vec{\eta}$, i.e., $G_{c} \in \Mcal_{3}(\Rbb)$, rather than
$\Mcal_{5}(\Rbb)$, and $\vec{q}_{c} \left( \nabla \Jcal \right) \in
\Rbb^{3}$, rather than $\Rbb^{5}$.

\section{Results from decoupling-pulse criteria + optimal control
  theory}
\label{sec:results-DPC+OCT}

Using the DPC+OCT protocol described in section \ref{sec:DPC+OCT} and
the DPs in fig.~\ref{fig:fields-DPC-Z2-Z} for the initial iterations of
all values of $\varepsilon_{0}$ considered, we sought to numerically
explore the space of controls satisfying $\vec{\eta}^{\, \rmr} = 0$ and
$\Delta = 0$ to improve control fidelity \emph{and} robustness to
$\varepsilon$-uncertainty for $Z_{\pi/2}$ and $Z_{\pi}$ operations,
compared to the original DPs. To a certain extent, it appears \emph{a
priori} that the minimization of $\Delta$ and $\vec{\eta}^{\, \rmr}$
might be competing control objectives. For example, compare the
gate-distance plots for OCT (figs.~\ref{fig:distance-OCT-Z2} and
\ref{fig:distance-OCT-Z}) to those for the DPs
(fig.~\ref{fig:distance-DPC-Z2-Z}) over the interval $0 \leq 
\varepsilon \leq 6$ (with a numerical resolution of 0.01 scaled
units). OCT for the design of unitary operations, as we have formulated
it in section \ref{sec:control-GrA}, seeks to minimize $\Delta$ for a
\emph{particular} value of $\varepsilon$, namely, the parameter estimate
$\varepsilon_{0}$. Because the system described by the Hamiltonian in
eq.~\eqref{eq:Hamiltonian} is controllable and the underlying control
landscape possesses a fortuitous structure for regular controls
\cite{Ho09a, Dominy11a}, a GrA achieves this objective quite efficiently
and successfully. However, as presented in section
\ref{sec:results-OCT}, these OCs are not inherently robust to
perturbations in $\varepsilon$, whereas controls satisfying
$\vec{\eta}^{\, \rmr} = 0$ are nearly optimal (with respect to $\Jcal$)
when $\varepsilon_{0} \approx 0$. This scenario illustrates the
potential balance that can exist between fidelity and robustness in
general. Overall, the controls that satisfy $\vec{\eta}^{\, \rmr} = 0$
are reasonably robust to perturbations in $\varepsilon$, provided that
$\varepsilon$ and/or $\Delta \varepsilon$ are within the perturbative
limit. Given that these two control objectives are potentially
competing, we use the hybrid optimization procedure developed in
section~\ref{sec:DPC+OCT} to suppress deviations from $\vec{\eta}^{\,
\rmr} = 0$, but not entirely eliminate them. In other words, convergence
of the DMORPH DPC+OCT algorithm occurs only when $\Jcal$ stops
decreasing, not when $\vec{\eta}^{\, \rmr}$ starts increasing.

\begin{figure}
\centering
\includegraphics[width=0.8\textwidth]{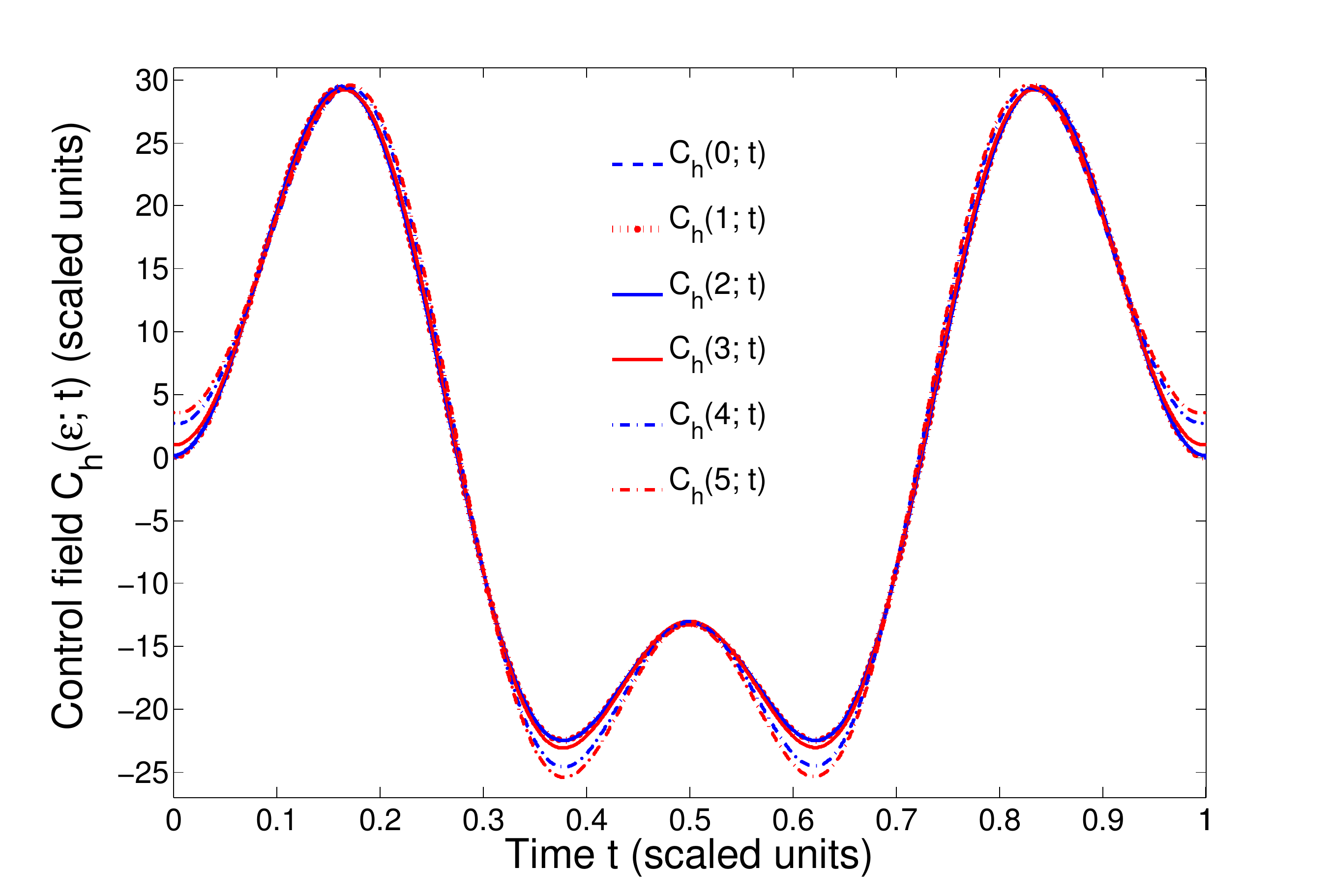}
\caption{(Color online) DPC+OCT fields $C_{\rmh}(\varepsilon_{0}; t)$
  producing a $Z_{\pi/2}$ operation, optimized using $C_{\rmd}(t)$ as the
  initial control for all estimates/values of $\varepsilon_{0}$ and $\tf
  = 1$ scaled unit. Although distinct, note that $C_{\rmh}(0; t)$,
  $C_{\rmh}(1; t)$, and $C_{\rmh}(2; t)$ appear nearly indistinguishable
  in this figure.}
\label{fig:fields-DPC+OCT-Z2}
\end{figure}

\begin{figure}
\centering
\includegraphics[width=0.8\textwidth]{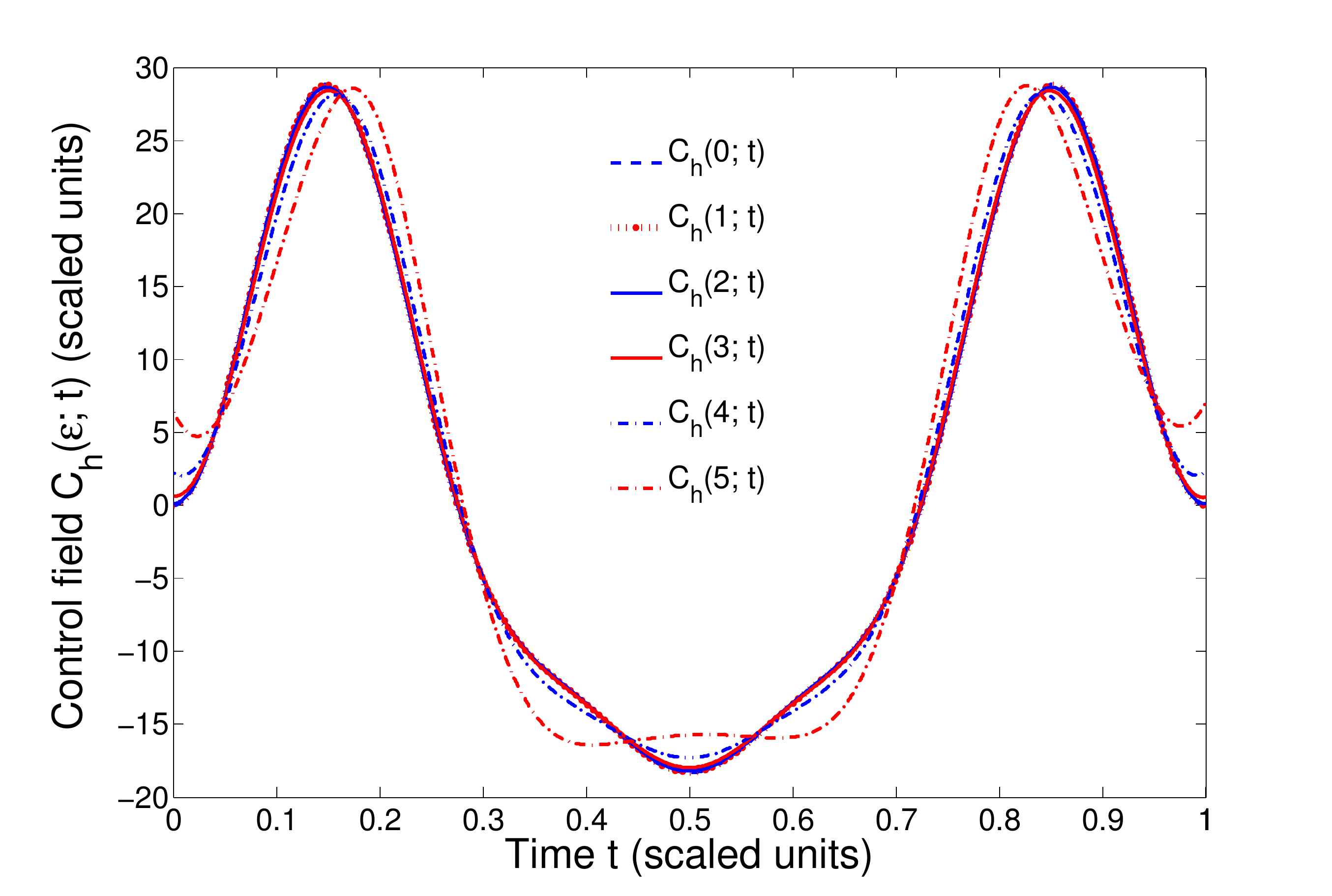}
\caption{(Color online) DPC+OCT fields $C_{\rmh}(\varepsilon_{0}; t)$
  producing a $Z_{\pi}$ operation, optimized using $C_{\rmd}(t)$ as the
  initial control for all estimates/values of $\varepsilon_{0}$ and $\tf
  = 1$ scaled unit. Although distinct, note that $C_{\rmh}(0; t)$,
  $C_{\rmh}(1; t)$, and $C_{\rmh}(2; t)$ appear nearly indistinguishable
  in this figure.}
\label{fig:fields-DPC+OCT-Z}
\end{figure}

\begin{figure}
\centering
\includegraphics[width=0.8\textwidth]{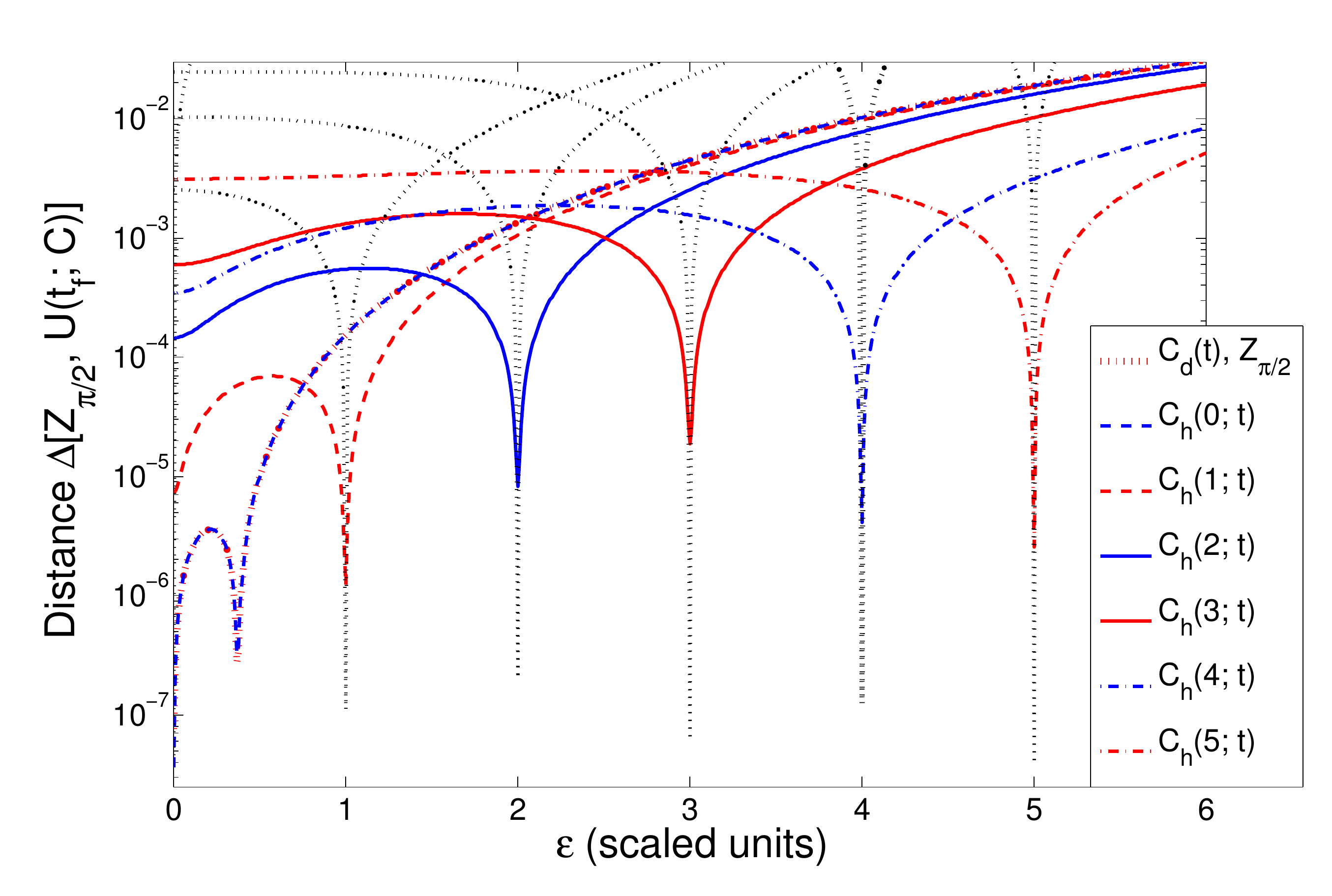}
\caption{(Color online) Distance for the $Z_{\pi/2}$ operation as a
  function of $\varepsilon$ (with a resolution of 0.01 scaled units) for
  control fields satisfying (a) $\vec{\eta}^{\, \rmr} = 0$ [red dashed
  line, which is very similar to that for $C_{\rmh}(0; t)$], (b)
  $\vec{\eta}^{\, \rmr} \approx 0$ and $\Delta \approx 0$, optimized
  with a \emph{specified} value of $\varepsilon_{0}$, and (c) results
  from $C_{\rmo}(\varepsilon_{0}; t)$ in section~\ref{sec:results-OCT}
  (black dashed lines, from fig.~\ref{fig:distance-OCT-Z2}).}
\label{fig:distance-DPC+OCT-Z2}
\end{figure}

\begin{figure}
\centering
\includegraphics[width=0.8\textwidth]{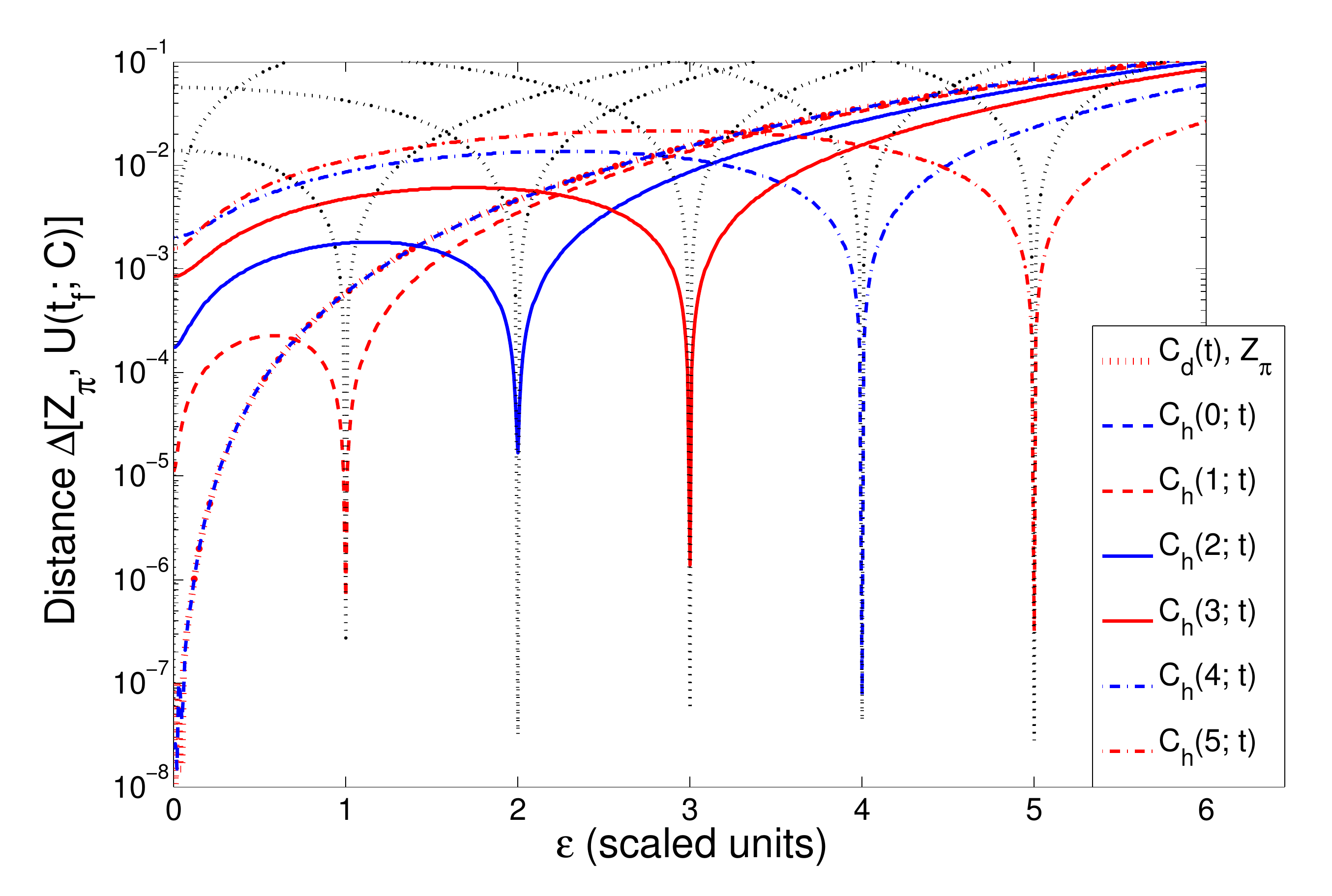}
\caption{(Color online) Distance for the $Z_{\pi}$ operation as a
  function of $\varepsilon$ (with a resolution of 0.01 scaled units) for
  control fields satisfying (a) $\vec{\eta}^{\, \rmr} = 0$ [red dashed
  line, which is very similar to that for $C_{\rmh}(0; t)$)], (b)
  $\vec{\eta}^{\, \rmr} \approx 0$ and $\Delta \approx 0$, optimized
  with a \emph{specified} value of $\varepsilon_{0}$, and (c) results
  from $C_{\rmo}(\varepsilon_{0}; t)$ in section~\ref{sec:results-OCT}
  (black dashed lines, from fig.~\ref{fig:distance-OCT-Z}).}
\label{fig:distance-DPC+OCT-Z}
\end{figure}

\begin{figure}
\centering
\includegraphics[width=0.8\textwidth]{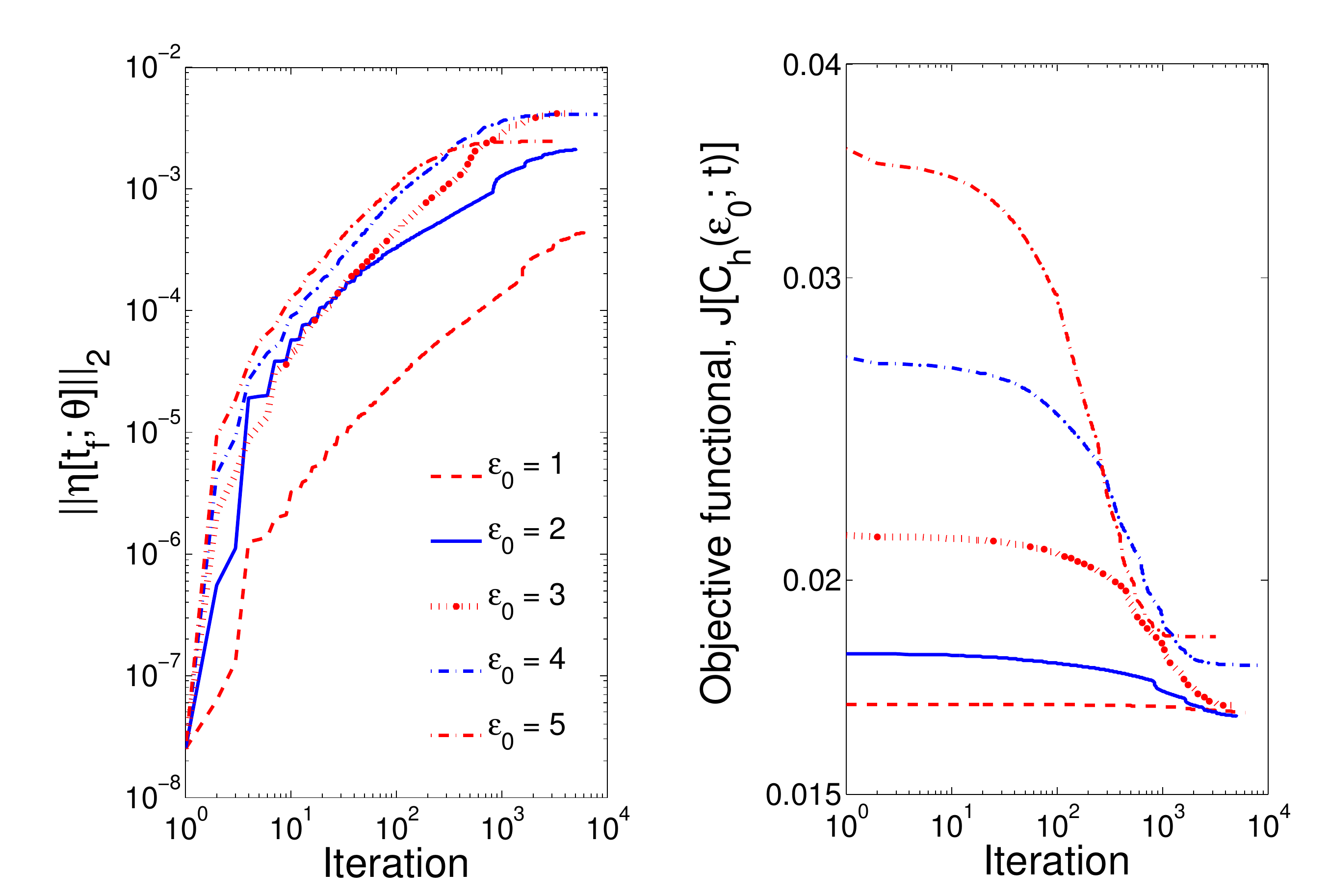}
\caption{(Color online) $\| \vec{\eta}^{\, \rmr} \|_{2}$ and objective
  functional $\Jcal$ for the $Z_{\pi/2}$ operation and each value of
  $\varepsilon_{0}$ considered, as a function of the number of DPC+OCT
  algorithm iterations.}
\label{fig:eta-J-DPC+OCT-Z2}
\end{figure}

\begin{figure}
\centering
\includegraphics[width=0.8\textwidth]{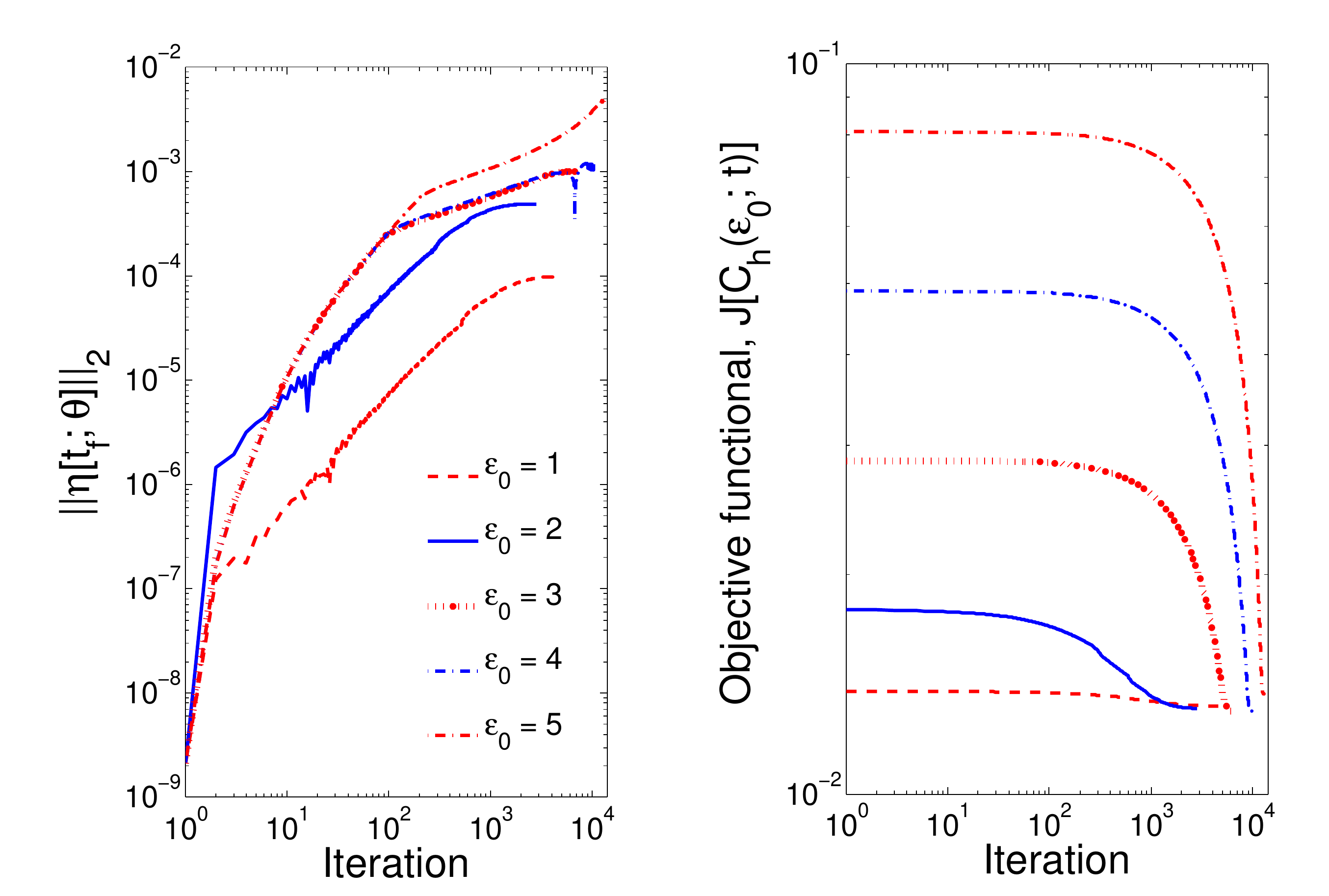}
\caption{(Color online) $\| \vec{\eta}^{\, \rmr} \|_{2}$ and objective
  functional $\Jcal$ for the $Z_{\pi}$ operation and each value of
  $\varepsilon_{0}$ considered, as a function of the number of DPC+OCT
  algorithm iterations.}
\label{fig:eta-J-DPC+OCT-Z}
\end{figure}

Because the formulation of OCT requires the specification of
$\varepsilon$, we consider $\varepsilon_{0} \in [0, 5]$. As before, the
final time for all controls was fixed at $\tf = 1$ scaled unit of time.
DPC+OCT fields for $Z_{\pi/2}$ and $Z_{\pi}$ are presented in
fig.~\ref{fig:fields-DPC+OCT-Z2} and fig.~\ref{fig:fields-DPC+OCT-Z}, 
respectively. These fields are denoted as $C_{\rmh}(\varepsilon_{0};
t)$, where the subscript ``h'' indicates the hybrid feature of this
QCP. Corresponding results for the gate distance $\Delta$ are presented
in figs.~\ref{fig:distance-DPC+OCT-Z2} and \ref{fig:distance-DPC+OCT-Z};
results for the vector-constraint norm $\| \vec{\eta}^{\, \rmr} \|_{2}$,
and objective functional $\Jcal$ are presented in
figs.~\ref{fig:eta-J-DPC+OCT-Z2} and \ref{fig:eta-J-DPC+OCT-Z}. In
addition, table \ref{tab:DPC+OCT-parameters} contains information about
some of the properties of the DPC+OCT fields. For both $Z_{\phi}$
operations, when $\varepsilon_{0} \geq 2$, we note that $\|
\vec{\eta}^{\, \rmr} \|_{2} > 10^{-4}$. However, since the robustness
criteria quantified by $\vec{\eta}^{\, \rmr}$ were obtained from a
perturbative analysis about $\varepsilon_{0} = 0$, it remains an open
question whether, \emph{a priori}, control fields satisfying
$\vec{\eta}^{\, \rmr} = 0$ for $\varepsilon_{0} > 0$ will be robust to
fluctuations about $\varepsilon_{0}$. However, as we present in this
section, control solutions obtained from the DPC+OCT protocol have some
desireable properties, even when $\varepsilon_{0} > 0$. It is
interesting to explore the results of the DPC+OCT protocol when the
distance is relatively large (i.e., $\Delta > 10^{-3}$) and the
sensitivity to changes in $\varepsilon$ is small, e.g., as illustrated
in fig.~\ref{fig:distance-DPC-Z2-Z}.

Despite their unique gate-distance dependence of these control solutions
on $\varepsilon$, as shown in figs.~\ref{fig:distance-DPC+OCT-Z2} and
\ref{fig:distance-DPC+OCT-Z} over the interval $0 \leq \varepsilon \leq
6$ (with a numerical resolution of 0.01 scaled units), the converged
DPC+OCT fields for $\varepsilon_{0} \in \{0, 1, 2, 3\}$ are very similar
to each other and the originated DP for each target unitary
operation. This supports the observation in
section~\ref{sec:results-OCT} that this simple system can effectively
discriminate between very similar control fields, i.e., as measured by
the gate distance $\Delta$ in eq.~\eqref{eq:distance}, the qubit system
is quite sensitive to these relatively small control-field
variations. For example, although $\max_{t} \left| C_{\rmh}(1; t) -
C_{\rmh}(2; t) \right| < 0.2$ scaled units of energy for both
operations, and the mean relative difference is approximately 1.5\% and
1.3\% for the $Z_{\pi/2}$ and $Z_{\pi}$ operations, respectively, the
corresponding gate distances (presented in
figs.~\ref{fig:distance-DPC+OCT-Z2} and \ref{fig:distance-DPC+OCT-Z}) do
not coincide significantly when $0 \leq \varepsilon \leq 3$.

Interestingly, the DPC+OCT control field for the $Z_{\pi}$ operation
when $\varepsilon_{0} = 5$ has some distinguishing features compared to
the fields for the other values of $\varepsilon_{0}$. Based on the
relatively large distance $\Delta$ of the corresponding $C_{\rmd}(t)$
control used for the initial iterate in the DPC+OCT protocol ($\Delta
= 6.75 \times 10^{-2}$), this value of $\varepsilon_{0}$ is not within
the perturbative limit of the analysis that produced the vector
constraint $\vec{\eta}^{\, \rmr} = 0$. With such a large distance at
$\varepsilon_{0} = 5$, the DPC+OCT routine improves the distance by
a factor larger than $10^{4}$ and simultaneously improves the robustness
$\Rcal_{\pi}$ for $4.5 \leq \varepsilon \leq 5.5$ by a factor larger
than 10, compared to the corresponding results for $C_{\rmd}(t)$
presented in table \ref{tab:DPC-parameters}.

Figures \ref{fig:distance-DPC+OCT-Z2} and \ref{fig:distance-DPC+OCT-Z}
compare the distance of $C_{\rmd}(t)$ and $C_{\rmh}(\varepsilon_{0}; t)$
control fields for both $Z_{\phi}$ operations. Compared to $C_{\rmd}(t)$,
all control fields $C_{\rmh}(\varepsilon_{0}; t)$ for $\varepsilon_{0}
\geq 0$ exhibit improved robustness to $\varepsilon$-uncertainties in an
interval around the nominal value $\varepsilon_{0}$ used in the DPC+OCT
algorithm. This result demonstrates the utility of combining so-called 
``pre-design'' methods, which are based on mathematically analyzing
general models [e.g., eq.~\eqref{eq:Hamiltonian-open1}], such as the DPC
developed by Pasini \emph{et al.} \cite{Pasini09a}, with numerical OCT
procedures and simple estimates of system parameters (e.g., estimates of
$\varepsilon_{0}$), especially when capabilities for shaping control
fields are available. By combining these QCPs, we have developed a form
of hybrid quantum control; estimates of system parameters can be
directly incorporated into simulations to generate improved quantum
operations for information processing and memory.

Figures \ref{fig:eta-J-DPC+OCT-Z2} and \ref{fig:eta-J-DPC+OCT-Z} compare
the vector-constraint norm $\| \vec{\eta}^{\, \rmr} \|_{2}$ and objective
functional $\Jcal$ as a function of the optimization iteration for both
$Z_{\phi}$ operations. Overall, $\| \vec{\eta}^{\, \rmr} \|_{2}$
increases as $\Jcal$ decreases, which is consistent with the notion of
minimizing $\| \vec{\eta}^{\, \rmr} \|_{2}$ and $\Jcal$ as potentially
competing control objectives. Even though components of $\nabla \Jcal$
that are parallel to all gradients $\nabla \eta_{i}$ are removed at each
iteration, $\| \vec{\eta}^{\, \rmr} \|_{2}$ increases during the
optimization for (at least) two reasons: (a) eq.~\eqref{eq:project}
removes components of $\nabla \eta_{i}$ (where elements $\eta_{i}$ are
\emph{nonlinear} functions of the control) using an iterative linear
projection method and (b) convergence of the DPC+OCT routine does not
depend on $\| \vec{\eta}^{\, \rmr} \|_{2}$.

\begin{figure}
\centering
\includegraphics[width=0.8\textwidth]{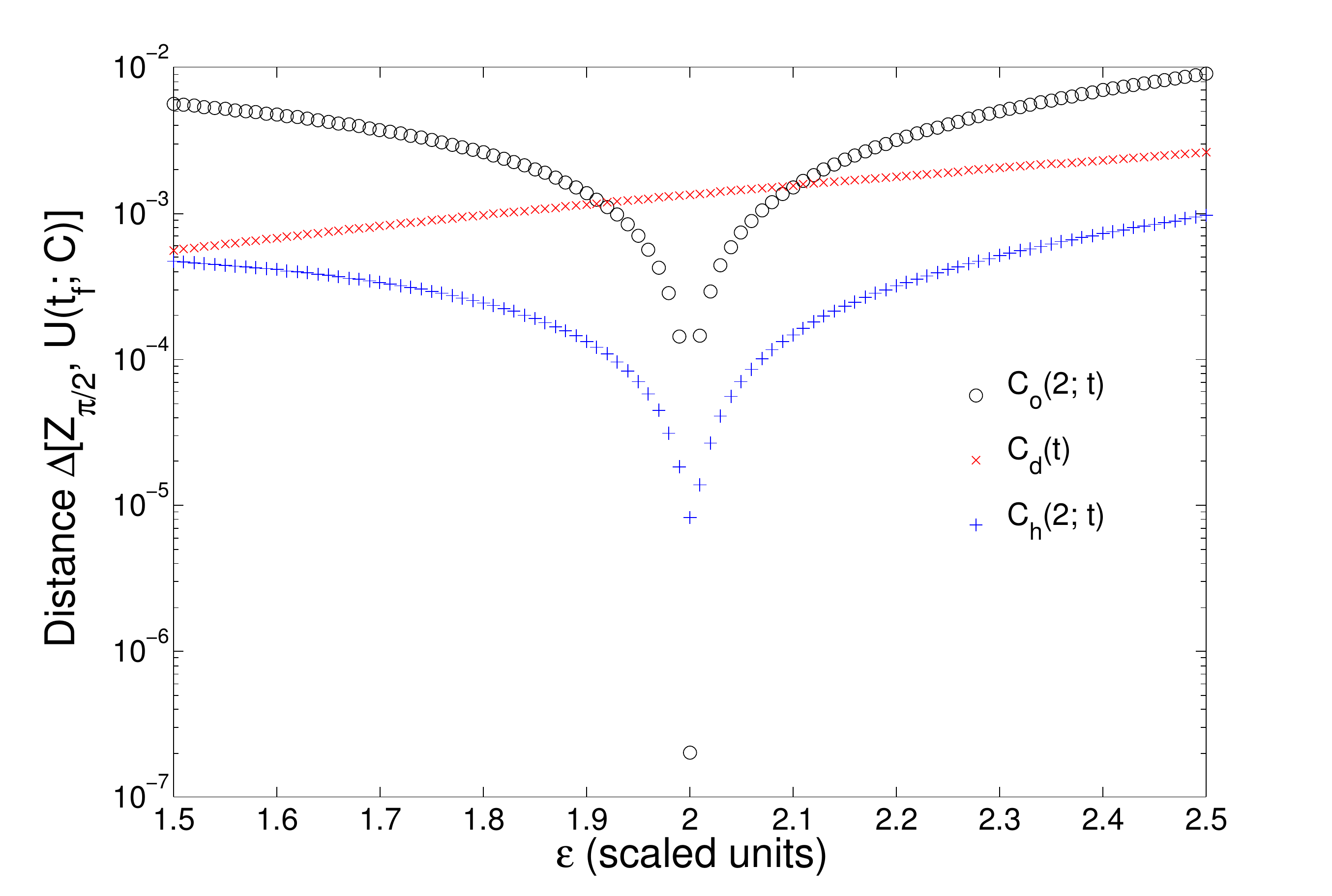}
\caption{(Color online) Distance for the $Z_{\pi/2}$ operation as a
  function of $\varepsilon$ for $C_{\rmo}(2; t)$, $C_{\rmd}(t)$, and
  $C_{\rmh}(2; t)$ controls in a unit interval centered at
  $\varepsilon_{0} = 2$ (with a resolution of 0.01 scaled units). Note
  that for $\varepsilon \geq 0.7$, $\Delta[Z_{\pi}, U(\tf; C_{\rmd})]
  \geq \Delta[Z_{\pi}, U(\tf; C_{\rmh})]$, i.e., $C_{\rmh}(1; t)$
  outperforms $C_{\rmd}(t)$.}
\label{fig:distance-OCT-DPC+OCT-Z2-epv-2}
\end{figure}

\begin{figure}
\centering
\includegraphics[width=0.8\textwidth]{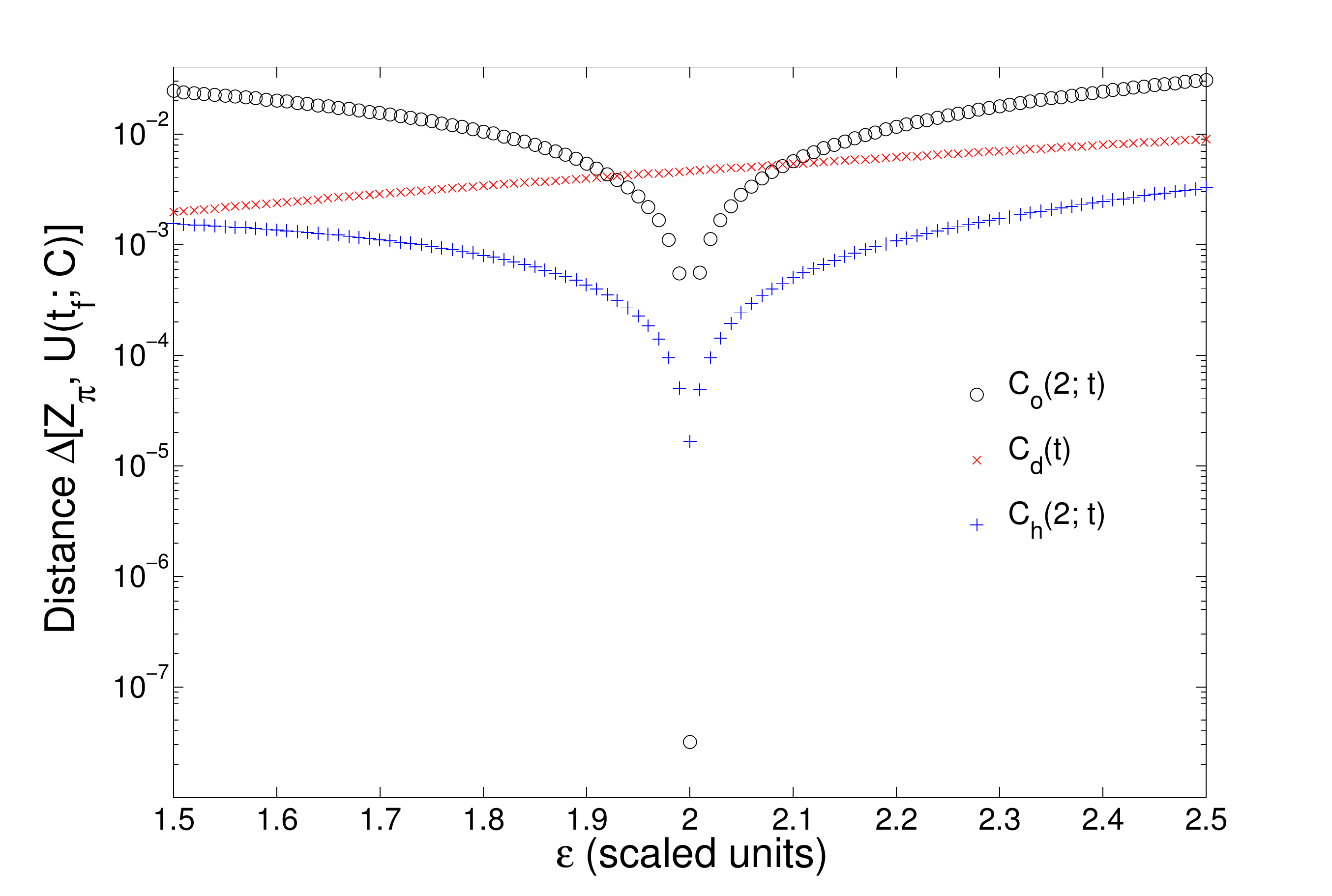}
\caption{(Color online) Distance for the $Z_{\pi}$ operation as a
  function of $\varepsilon$ for $C_{\rmo}(2; t)$, $C_{\rmd}(t)$, and
  $C_{\rmh}(2; t)$ controls in a unit interval centered at
  $\varepsilon_{0} = 2$ (with a resolution of 0.01 scaled units). Note
  that for $\varepsilon \geq 1.5$, $\Delta[Z_{\pi}, U(\tf; C_{\rmd})]
  \geq \Delta[Z_{\pi}, U(\tf; C_{\rmh})]$, i.e., $C_{\rmh}(1; t)$
  outperforms $C_{\rmd}(t)$.}
\label{fig:distance-OCT-DPC+OCT-Z-epv-2}
\end{figure}

To aid in the comparative analysis of results from OCT, DPC, and the
DPC+OCT procedures, OC gate-distance data from
figs.~\ref{fig:distance-OCT-Z2} and \ref{fig:distance-OCT-Z} are also
presented in figs.~\ref{fig:distance-DPC+OCT-Z2} and
\ref{fig:distance-DPC+OCT-Z}, respectively. Although 
the OC fields $C_{\rmo}(\varepsilon_{0}; t)$ all outperform the
DPC+OCT fields $C_{\rmh}(\varepsilon_{0}; t)$ at $\varepsilon_{0}$, these
OCs do not have the robustness of the DPs or DPC+OCT fields. To
emphasize this feature, figs.~\ref{fig:distance-OCT-DPC+OCT-Z2-epv-2}
and \ref{fig:distance-OCT-DPC+OCT-Z-epv-2} present $Z_{\pi/2}$ and
$Z_{\pi}$ gate distances for $C_{\rmo}(2; t)$, $C_{\rmd}(t)$, and
$C_{\rmh}(2; t)$ controls over a unit interval centered at
$\varepsilon_{0} = 2$. For both gates, $C_{\rmo}(2; t)$ is very
sensitive to variations in $\varepsilon$, e.g., when $\varepsilon$
changes from 2 to $2 \pm 0.01$ (a change corresponding to approximately
$1.3 \times 10^{-5}$ T for the GaAs DQD example), the $Z_{\pi/2}$ gate
distance increases (approximately) from $10^{-7}$ to $10^{-4}$,
while the $Z_{\pi}$ gate distance increases from $10^{-8}$ to $10^{-4}$,
approaching the fault-tolerant threshold. However, for $C_{\rmh}(2; t)$
for both gates, as $\varepsilon$ varies from $\varepsilon_{0}$, the
increase in gate distance is much more gradual. Figures
\ref{fig:distance-OCT-DPC+OCT-Z2-epv-2} and
\ref{fig:distance-OCT-DPC+OCT-Z-epv-2} contain some useful 
information to help understand the benefit of the hybrid DPC+OCT
protocol. By combining DPC, OCT, a DP that satisfies $\vec{\eta}^{\,
\rmr} = 0$ for the initial GrA iteration, and an estimate of the value
of $\varepsilon_{0}$, the gate distance is decreased compared to the
gate distance of the original DP for the entire unit interval centered
at $\varepsilon_{0}$. Depending on the uncertainty magnitude of
$\varepsilon$, this benefit could yield a potentially substantial
decrease in the required concatenation/encoding resources necessary for
QECCs, which depend on gate errors.

\begin{figure}
\centering
\includegraphics[width=0.8\textwidth]{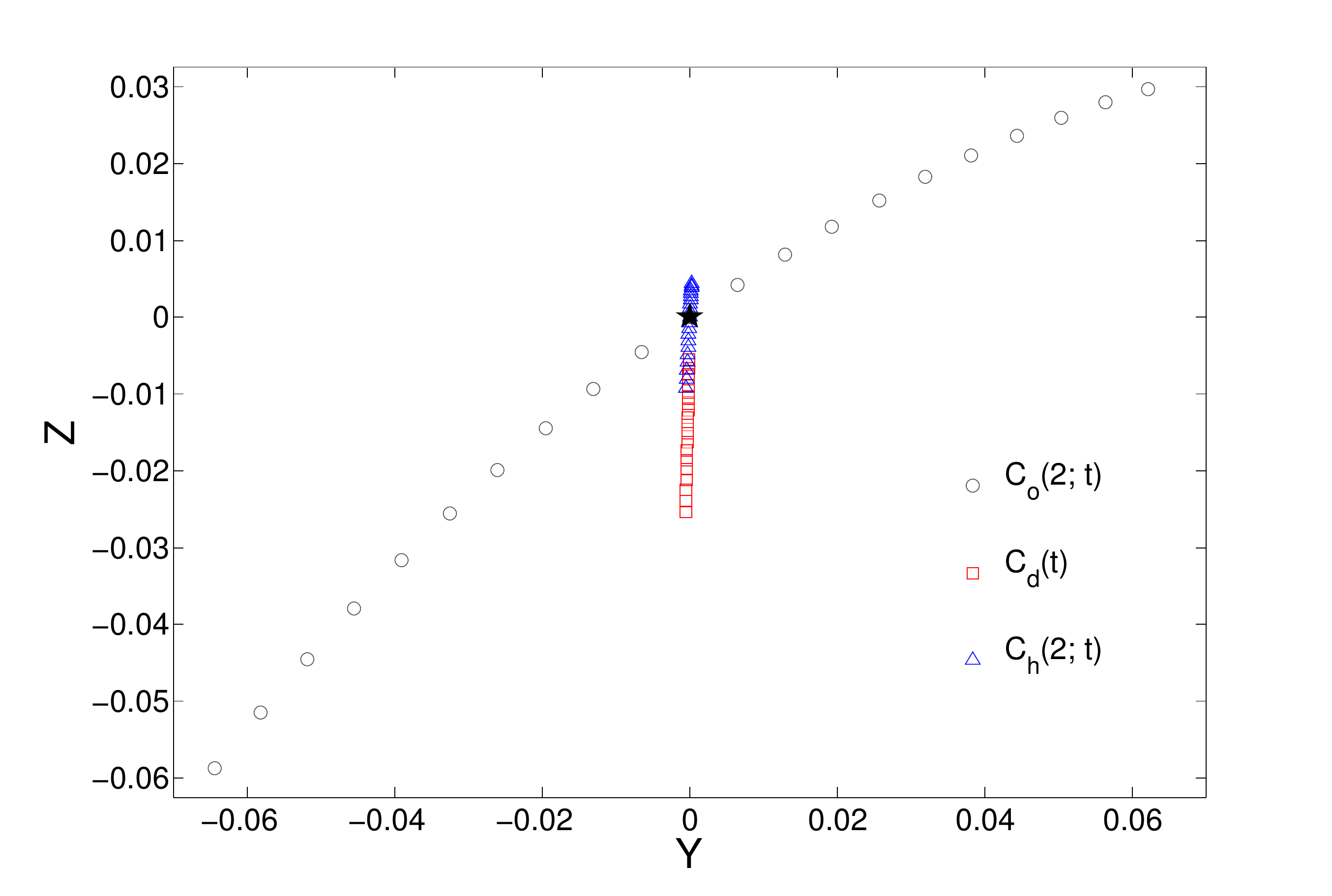}
\caption{(Color online) Final state in the Bloch vector coordinates $y$
  and $z$ for the $Z_{\pi}$ operation, implemented with the
  corresponding controls $C_{\rmo}(2; t)$, $C_{\rmd}(t)$, and
  $C_{\rmh}(2; t)$ applied to an ensemble of systems described by the
  Hamiltonian in eq.~\eqref{eq:Hamiltonian} and the interval $1.5 \leq
  \varepsilon \leq 2.5$ (distributed over 20 equal increments of 0.05
  scaled units). The target state is $|\sigma_{x}^{-}\rangle =   Z_{\pi} 
  |\sigma_{x}^{+}\rangle$. Because $-1 \leq x < -0.995$ for all final
  states, it is not included in this figure.}
\label{fig:BlochVector-OCT-DPC+OCT_Z-X_eps-2}
\end{figure}

As a final illustrative example, consider the $Z_{\pi}$ operation
applied to the initial state $|\sigma_{x}^{+}\rangle$, implemented with
the corresponding controls $C_{\rmo}(2; t)$, $C_{\rmd}(t)$, and
$C_{\rmh}(2; t)$, which are applied to an ensemble of systems described
by the Hamiltonian in eq.~\eqref{eq:Hamiltonian} and the interval $1.5
\leq \varepsilon \leq 2.5$ (numerically distributed over 20 equal
increments of 0.05 scaled units). The target state is
$|\sigma_{x}^{-}\rangle = Z_{\pi} |\sigma_{x}^{+}\rangle$, where
$\sigma_{x} |\sigma_{x}^{\pm}\rangle = \pm |\sigma_{x}^{\pm}\rangle$,
and the ensemble of final states for a given control is denoted by
$\{|\psi_{i}\rangle\}$. This state-based example clarifies the gate
improvement obtained from $C_{\rmh}(2; t)$, compared to $C_{\rmo}(2; t)$
and $C_{\rmd}(t)$. To quantify the fidelity of the controls, we use the
Uhlmann state fidelity for pure states
\cite{Uhlmann76a, Jozsa94a}:
\begin{equation}
\label{eq:Uhlmann}
\Fcal_{\rmu} \left( |\psi_{1}\rangle, |\psi_{2}\rangle \right) :=
\left| \langle \psi_{1} | \psi_{2} \rangle \right|, 
\end{equation} 
where $|\psi_{1}\rangle$ and $|\psi_{2}\rangle$ are normalized vectors
in $\Hcal$. For a given control, we denote the resulting minimum,
maximum, and average state fidelity, and the standard deviation of the
fidelity of the ensemble as $\min_{|\psi_{i}\rangle} \Fcal_{\rmu}$,
$\max_{|\psi_{i}\rangle} \Fcal_{\rmu}$, $\bar{\Fcal}_{\rmu}$, and
$\sigma_{\Fcal_{\rmu}}$, respectively, which are presented in table
\ref{tab:state-fidelity} for $C_{\rmo}(2; t)$, $C_{\rmd}(t)$, and
$C_{\rmh}(2; t)$. Comparing the respective quantities, $C_{\rmh}(2; t)$
has the largest average and minimum fidelity and the smallest standard
deviation of fidelity of the ensemble (by nearly a factor of 10). The
final-time ensembles are also illustrated in
fig.~\ref{fig:BlochVector-OCT-DPC+OCT_Z-X_eps-2}, which contains a
plot of resulting final states for each control, along with the target
state $|\sigma_{x}^{-}\rangle$, all in the Bloch vector coordinates $y$
and $z$. Because $-1 \leq x < -0.995$ for all final states, it is
not included in this figure. Unlike the Bloch vector components
corresponding to the final states produced from $C_{\rmo}(2; t)$ and
$C_{\rmd}(t)$, the Bloch vector components produced from $C_{\rmh}(2;
t)$ are tightly distributed around the target state, with most of the
error distributed uniformly along the $z$-axis, centered at the target
state $|\sigma_{x}^{-}\rangle$. Very similar results are obtained for the
$Z_{\pi/2}$ operation implemented with $C_{\rmo}(2; t)$, $C_{\rmd}(t)$,
and $C_{\rmh}(2; t)$, applied to an ensemble of systems where $1.5 \leq
\varepsilon \leq 2.5$. 

\begin{table}
\begin{tabular}{@{}|c|c|c|c|c|c|c|}
\multicolumn{7}{c}{\textbf{Target operation:} $\bm{Z_{\pi/2}}$} \\
\hline
$\varepsilon_{0}$ & 0 & 1 & 2 & 3 & 4 & 5 \\
\hline\hline
$\max|C_{\rmh}|$ & 29.5 & 29.5 & 29.3 & 29.2 & 29.4 & 29.6 \\
\hline
$\theta(\tf; C_{\rmh})$ & $\pi/2$ & $\pi/2$ & 1.5704 & 1.5690 & 1.5695 &
1.5792 \\
\hline
$\Phi[C_{\rmh}]$ & 335.5 & 334.6 & 332.8 & 337.4 & 356.5 & 370.5 \\
\hline
\ $\Delta(Z_{\pi/2}, \Utf)$ \ & \ $3.65 \times 10^{-8}$ \ &
\ $1.21 \times 10^{-6}$ \ & \ $8.23 \times 10^{-6}$ \ &
\ $1.86 \times 10^{-5}$ \ & \ $4.08 \times 10^{-6}$ \ & 
\ $2.55 \times 10^{-6}$ \ \\
\hline
$\| \vec{\eta}^{\, \rmr} (\tf; C_{\rmh}) \|_{2}$ & $3.21 \times 10^{-8}$ &
$4.74 \times 10^{-4}$ & $2.13 \times 10^{-3}$ & $4.21 \times 10^{-3}$ &
$4.12 \times 10^{-3}$ & $2.45 \times 10^{-3}$ \\
\hline
\ $\Rcal_{\pi/2}[C_{\rmh}, \varepsilon_{0}, 0.5]$ \ & \ $2.84 \times
10^{-6}$ \ &
\ $9.47 \times 10^{-5}$ \ & \ $3.55 \times 10^{-4}$ \ &
\ $6.46 \times 10^{-4}$ \ & \ $5.72 \times 10^{-4}$ \ & 
\ $9.20 \times 10^{-4}$ \ \\
\hline
\multicolumn{7}{c}{\textbf{Target operation:} $\bm{Z_{\pi}}$} \\
\hline
$\varepsilon_{0}$ & 0 & 1 & 2 & 3 & 4 & 5 \\
\hline\hline
$\max|C_{\rmh}|$ & 28.8 & 28.8 & 28.6 & 28.4 & 28.4 & 28.9 \\
\hline
$\theta(\tf; C_{\rmh})$ & $\pi$ & $\pi$ & 3.1411 & 3.1392 & 3.1343 &
3.1320 \\
\hline
$\Phi[C_{\rmh}]$ & 264.8 & 264.0 & 261.8 & 259.1 & 260.0 & 280.7 \\
\hline
\ $\Delta(Z_{\pi}, \Utf)$ \ & \ $2.36 \times 10^{-8}$ \ &
\ $6.57 \times 10^{-7}$ \ & \ $1.67 \times 10^{-5}$ \ &
\ $1.74 \times 10^{-5}$ \ & \ $2.79 \times 10^{-8}$ \ & 
\ $1.10 \times 10^{-6}$ \ \\
\hline
$\| \vec{\eta}^{\, \rmr} (\tf; C_{\rmh}) \|_{2}$ & $3.55 \times 10^{-9}$ &
$9.93 \times 10^{-5}$ & $4.90 \times 10^{-4}$ & $1.10 \times 10^{-3}$ &
$6.49 \times 10^{-3}$ & $2.10 \times 10^{-2}$ \\
\hline
\ $\Rcal_{\phi}[C_{\rmh}, \varepsilon_{0}, 0.5]$ \ & \ $1.83 \times
10^{-5}$ \ &
\ $3.11 \times 10^{-4}$ \ & \ $1.18 \times 10^{-3}$ \ &
\ $2.59 \times 10^{-3}$ \ & \ $4.23 \times 10^{-3}$ \ & 
\ $4.92 \times 10^{-3}$ \ \\
\hline
\end{tabular}
\caption{Performance of the DPC+OCT controls $C_{\rmh}(\epsilon_{0};
  t)$ for one-qubit $Z_{\phi}$ operations. Here, $\max|C_{\rmh}|$,
  $\theta$, $\Phi[C] := \int_{0}^{\tf} \! C^{2}(\varepsilon_{0}; t)
  \rmd t$, $\Delta$, $\| \vec{\eta}^{\, \rmr} \|_{2}$, and $\Rcal_{\phi}$ are
  the maximum control-field amplitude, angle of controlled $z$-axis
  rotation, control-field fluence, gate distance, constraint vector
  norm, and gate robustness, respectively, in the corresponding
  scaled units described in section~\ref{ssec:units}.}
\label{tab:DPC+OCT-parameters}
\end{table}

\begin{table}
\begin{tabular}{@{}|c|c|c|c|c|}
\multicolumn{5}{c}{\textbf{Target state:}
$\bm{|\sigma_{x}^{-}\rangle = Z_{\pi} |\sigma_{x}^{+}\rangle}$} \\
\hline
 & \ \ $\displaystyle{\min_{|\psi_{i}\rangle} \Fcal_{\rmu} \left[
    |\sigma_{x}^{-}\rangle, |\psi_{i}\rangle \right]}$ \ \ & \ \
$\displaystyle{\max_{|\psi_{i}\rangle} \Fcal_{\rmu} \left[
    |\sigma_{x}^{-}\rangle, |\psi_{i}\rangle \right]}$ \ \ &
$\bar{\Fcal}_{\rmu}$ & $\sigma_{\Fcal_{\rmu}}$ \\
\hline
\ \ $C_{\rmo}(2; t)$ \ \ & \ \ 0.996197 \ \ & \ \ 1.0 \ \ & \ \ 0.998871
\ \ & \ \ $1.082 \times 10^{-3}$ \ \ \\
\hline
$C_{\rmd}(t)$ & 0.999678 & \ \ 0.999985 \ \ & 0.999884 & $9.398 \times
10^{-5}$ \\
\hline
$C_{\rmh}(2; t)$ & 0.999958 & 1.0 & 0.999991 & $1.129 \times 10^{-5}$
\\
\hline
\end{tabular}
\caption{Properties of state fidelity for the transition
  $|\sigma_{x}^{-}\rangle = Z_{\pi} |\sigma_{x}^{+}\rangle$ driven by
  controls $C_{\rmo}(2; t)$, $C_{\rmd}(t)$, and $C_{\rmh}(2; t)$. Here,
  $\min_{|\psi_{i}\rangle} \Fcal_{\rmu}$, $\max_{|\psi_{i}\rangle}
  \Fcal_{\rmu}$, $\bar{\Fcal}_{\rmu}$, and $\sigma_{\Fcal_{\rmu}}$
  denote the minimum, maximum, and average fidelity, and the standard
  deviation of the fidelity, respectively, of the ensemble of final
  states $\{|\psi_{i}\rangle\}$ compared to the target state
  $|\sigma_{x}^{-}\rangle$.} 
\label{tab:state-fidelity}
\end{table}

\section{Conclusions and future directions}
\label{sec:conclusion}

Combining OCT with the DPC established by Pasini \emph{et al.}
introduces improvements to the control of quantum systems for 
information processing. Given a reasonable characterization of the
angular frequency $\varepsilon / \hbar$ of a persistent, but somewhat
uncertain rotation about the $x$-axis, a near-optimal fidelity can be
achieved for $\varepsilon \geq 0$. Furthermore, the resulting DPC+OCT
controls exhibit improved robustness to uncertainty in $\varepsilon$,
compared to the original DPs. The systematic integration of general DPC
and control-field shaping methods from OCT, therefore, promises
considerable improvement over DPC or OCT strategies alone. We have
provided a quantitative illustration for a logical qubit based on a DQD
system, with continuous controls that possess reasonable magnitudes
\cite{Nielsen10a}, based on the scaled-to-SI unit mapping.

We are currently investigating the benefits of these DPC+OCT $\pi/2$-
and $\pi$-pulses for memory and information processing in the presence
of a decohering spin bath. It will be useful to determine how these
pulses extend spin-echoes and improve general DD and
dynamically-corrected gate pulse sequences, such as those described in
refs.~\cite{Liu07a, Uhrig08a, Khodjasteh05a, Khodjasteh07a,
Khodjasteh09a, Khodjasteh09b, Khodjasteh10a, Pasini11a}. Future work
involves an exploration of unitary control sensitivity to fluctuations
in the control field itself (e.g., control noise). \emph{Post facto}
analysis of both the OCT and DPC+OCT results presented in this article
for the general qubit model suggests that these fluctuations may
contribute just as significantly to gate errors as corresponding
system and environment fluctuations. However, our optimization criteria
does not include robustness to control-field noise; such robustness may
be sacrificed in favor of the actual criteria. Given the ubiquity of
noise in classical controls and quantum-mechanical systems, constructing
controls and systems that are robust to their own noise is crucial for
practical fault-tolerant QC.

Extensions of the original analysis by Pasini \emph{et al.} are also
being considered. We are interested in generalizing their results to
include (a) arbitrary angle rotations and axes, (b) closed-system
perturbative expansions about any value of $\varepsilon$, rather than
only $\varepsilon = 0$, and (c) $\varepsilon$ as a stochastic
time-dependent variable/operator, which is relevant to previous research
on decoherence control, e.g., \cite{Young10a}. In addition, direct
minimization of $\Rcal_{\phi}$ [eq.~\eqref{eq:robust-distance}] or
\begin{equation}
\Lcal[C] := \int_{\varepsilon_{1}}^{\varepsilon_{2}} \! \left(
c_{1} \left\| \frac{\rmd \Utf(C)}{\rmd \varepsilon} \right\|_{\rHS} + c_{2}
\left\| \frac{\rmd^{2} \Utf(C)}{\rmd \varepsilon^{2}} \right\|_{\rHS}
\right) \rmd \varepsilon
\end{equation}
over $\Jcal^{-1}(0)$, where $c_{1}$, $c_{2} \in \Rbb$ weight the
relative significance of the two norms, are purely OCT means to improve
robustness to variations in $\varepsilon$ about any fixed interval
$\left[ \varepsilon_{1}, \varepsilon_{2} \right]$, which we are also
investigating.

\section*{Acknowledgements}

MDG thanks Paul T.~Boggs (SNL-CA), and Robert L.~Kosut (SC Solutions,
Inc.) for illuminating discussions on control and nonlinear
optimization. MDG and WMW thank Stefano Pasini and G\"{o}tz S. Uhrig
(Technische Universit\"{a}t Dortmund) for useful discussions regarding
ref.~\cite{Pasini09a}. This work was supported by the Laboratory
Directed Research and Development program at Sandia National
Laboratories. Sandia is a multi-program laboratory managed and operated
by Sandia Corporation, a wholly owned subsidiary of Lockheed Martin
Corporation, for the United States Department of Energy's National
Nuclear Security Administrationy under contract DE-AC04-94AL85000.

\bibliography{biblio}
\bibliographystyle{apsrev4-1}

\newpage
\section*{Figures}

\end{document}